\documentclass[12pt, draftclsnofoot, onecolumn]{IEEEtran}

\ifCLASSINFOpdf
\else
\fi
\usepackage{graphicx}
\usepackage{epstopdf}
\usepackage{stfloats}
\usepackage{bm}
\usepackage[cmex10]{amsmath}
\usepackage{algorithmic}
\usepackage{algorithm}
\usepackage{array}

\usepackage{mdwmath}
\usepackage{mdwtab}
\usepackage{multirow}

\usepackage{amsfonts}

\usepackage{amssymb}

\usepackage{color}

\begin{document}
	%
	\title{ Multi-Cell Interference Exploitation: A New Dimension in Cell Coordination }

	\author{Zhongxiang Wei,~\IEEEmembership{Member,~IEEE,}
		Christos Masouros,~\IEEEmembership{Senior Member,~IEEE,}
		Kai-Kit Wong,~\IEEEmembership{Fellow,~IEEE,}
		Xin Kang,~\IEEEmembership{Member,~IEEE}
		\thanks{Zhongxiang Wei, Christos Masouros, and Kai-Kit Wong are with the Department of Electronic and Electrical Engineering at the University College London, London, UK. Email: \{zhongxiang.wei, c.masouros, kai-kit.wong\}@ucl.ac.uk}
		\thanks{Xin Kang is with Shield Lab., Huawei Singapore Research Center, Singapore. Email: kang.xin@huawei.com}
		\thanks{This work was supported by the Engineering and Physical Sciences Research Council, UK, under project EP/R007934/1.} 
		
	}

	\maketitle

	\begin{abstract}
		In this paper, we propose a series of novel coordination schemes for multi-cell downlink communication. Starting from full base station (BS) coordination, we first propose a fully-coordinated scheme to exploit beneficial effects of both inter-cell and intra-cell interference, based on sharing both channel state information (CSI) and data among the BSs. 
		To reduce the  coordination overhead, we then propose a partially-coordinated scheme where only intra-cell interference is designed to be constructive while inter-cell is jointly suppressed by the coordinated BSs. Accordingly, the coordination only involves CSI exchange and the need for sharing data is eliminated. 
		To further reduce the coordination overhead, a third scheme is proposed, which only requires the knowledge of statistical inter-cell channels, at the cost of a slight increase on the transmission power. 
		For all the proposed schemes, imperfect CSI is considered. We minimize the total transmission power in terms of probabilistic and deterministic optimizations. Explicitly, the former statistically satisfies the users' signal-to-interference-plus-noise ratio (SINR) while the latter guarantees the SINR requirements in the worst case CSI uncertainties. 
		Simulation verifies that our schemes consume much lower power compared to the existing benchmarks, i.e., coordinated multi-point (CoMP) and coordinated-beamforming (CBF) systems, opening a new dimension on multi-cell coordination.

	\end{abstract}
	

	%
	\IEEEpeerreviewmaketitle

	\section{Introduction}
	
	Inter-cell and intra-cell interference  are two key challenges in the coming heterogeneous and dense networks, which  have been considered as fundamental limiting factors of the system performance for next wireless communications \cite{Christos2013Known}. In the past decades, several techniques have been proposed to enhance system performance in the presence of inter-cell and intra-cell interference. Among them, multi-cell coordination has attracted much attention, where multiple base stations (BSs) collaborate with others for joint signal processing, based on the shared channel state information (CSI) and/or intended transmitted data \cite{Lee2012Coordinated}. Depending on the coordination level among the BSs, multi-cell coordination system can be classified into two categories: coordinated-beamforming (CBF) and coordinated multi-point system (CoMP), and each of them is suited for different scenarios and system configurations.

	In the CBF  scheme, a user is supported by a single BS and each BS only needs to encode or decode the signal to/from the users within its cell. Hence, the impairment caused by the inter-cell interference is suppressed by cooperative beamforming, and the beamforming at each BS is computed as a function of globally shared CSI, obtained from the users via feedback channels  \cite{Ko2012Effective}. Since the BSs jointly optimize their beams to suppress the inter-cell gains, the CBF scheme is also named as interference coordination. 
	In this spirit, several designs have been conducted to optimize system performance, e.g., throughput maximization \cite{ Ko2012Effective}-\cite{Park2012New}  and power minimization \cite{Tolli2011Decentralized} \cite{Kim2013Power}.
	Downlink throughput  \cite{ Ko2012Effective}  \cite{Venturino2010Coordinated}    \cite{Huh2011Multi} \cite{Simsek2015Learning} \cite{Park2012New} and uplink throughput maximization \cite{Nguyen2014Sum} \cite{Wu2011Sum} were investigated by transmit and receive beamforming, where BSs only transmit or decode the signal to/from the users within its cell, and inter-cell interference was mitigated as noise. 
	Most relevant to the approaches considered in this paper, power minimization problem was investigated in  \cite{Tolli2011Decentralized} and  \cite{Kim2013Power}. The authors in  \cite{Tolli2011Decentralized} minimized power beamformer for a CBF system, where the minimum  beamformers can be obtained locally at each BS relying on limited backhaul information exchange between BSs.
	The authors in \cite{Kim2013Power} proposed a power efficient transceiver design for multi-cell coordination in cognitive radio networks, where secondary BSs are coordinated to serve secondary users while imposing limited interference for the primary users. 
	In summary, the beamforming design by the CBF scheme typically strikes the trade-off between eliminating the inter-cell interference and maximizing the signal-to-interference-plus-noise ratio (SINR) to/from the user within the cell of interest, which requires a modest amount of coordination overhead.
	
	
	
	With a higher level of coordination, the performance of multi-cell systems can be improved by CoMP scheme if BSs are linked by high-capacity delay-free links. Explicitly, the coordinated BSs can share not only CSI, but also the data to be transmitted. This is typically facilitated through a high bandwidth backhaul network implemented by optical fiber or millimeter wave links \cite{Lee2012Coordinated}.
	Since the intended transmitted data can be available at all the coordinate BSs, CoMP system transforms the multi-cell system into a multi-user multiple-input and multiple-output (MIMO) for which all links (including interfering ones) are exploited to carry useful data \cite{David2010Multi}. 
	Hence, CoMP is also refereed to network-level virtual  MIMO, which coordinates the simultaneous information transmissions from multiple BSs to the users, at the expense of increased coordination overhead for sharing the intended transmitted data.
	Based on the principle, the authors in \cite{Wang2017Joint} considered the joint optimization of user association, subchannel allocation, and power allocation in a heterogeneous CoMP system, where each user is associated with one macro BS in the center and multiple separated micro BSs.
	The authors in \cite{Ali2018Downlink} applied the CoMP into non-orthogonal multiple access systems, where CoMP transmission is to serve users experiencing severe inter-cell interference. 
	From an information theoretic point of view, CBF scheme can be interpreted as a interference channel while CoMP scheme can be interpreted as a broadcast channel, and the latter generally achieves better performance over the former scheme by exploiting inter-cell channel, at the cost of high overhead.

	It is worth noting that, regardless of CBF or CoMP schemes, intra-cell interference (multi-user interference) still needs to be strictly mitigated. Nevertheless, in view of the recently introduced concept of constructive interference (CI) exploitation, there is scope to exploit intra-cell and inter-cell interference as a source of useful signal.
	As a consequence, system performance can be further improved.
	CI was firstly introduced by \cite{Christos2007A} in code division multiple access systems. Then  \cite{Christos2011Correlation} further proposed that all the interference can be constructive by designing precoding in symbol level \cite{ Alodeh2016Energy} \cite{Spano2018Symbol}. 
	Recently, the concept of CI was applied into beamforming optimization \cite{Masouros2014Vector} \cite{Masouros2015Exploiting}, cognitive radio \cite{Law2017Transmit}, large-scale MIMO \cite{Amadori2017Large}, multiuser multi-input and single-output (MISO) \cite{Li2018Interference}, wireless power transfer systems \cite{Timotheou2016Exploiting}, constant envelop systems \cite{Amadori2017Constant} and physical layer security-aware systems \cite{Khandaker2018Constructive}. However, it should be pointed that all the aforementioned research only focused on single-cell systems, which can not be applied to multi-cell coordination systems. Besides, considering imperfect CSI in practice, the presence of CSI error makes the CI design in multi-cell systems more challenging.

	
	Motivated by the aforementioned open challenges, in this paper, we investigate different multi-cell coordination schemes to fully/partially utilize inter-cell and intra-cell interference, in the presence of imperfect or statistical CSI. Our contributions are summarized as follows:
	
	\begin{enumerate}[]
		\item We propose a series of schemes to fully/partially utilize inter-cell and intra-cell interference for multi-cell coordination systems, which require different levels of overhead for coordination. Firstly, a fully-coordinated CI (Full-CI) scheme is proposed to utilize both inter-cell and intra-cell interference as beneficial elements, on the basis of CSI and data being shared among the coordinated BSs. 
		
		\item Then, a partially-coordinated CI (Partial-CI) scheme is proposed to utilize intra-cell interference while suppressing inter-cell interference by joint precoding design.
		Since only CSI needs to be shared among BSs, the coordination overhead is significantly reduced compared to the Full-CI scheme.

		\item At last, a statistical inter-channel based CI (Stat-CI)  scheme is proposed to further reduce the coordination overhead, where BSs only need to know statistical CSI from the adjacent BSs for inter-cell interference suppressing while keeping intra-cell interference constructive.
		
		\item In all the above designs, we exploit robust precoding design to accommodate scenarios with imperfect CSI. In the presence of CSI errors,  we investigate the total power minimization problems in probabilistic and deterministic manners, respectively. Explicitly, multiple users' SINR requirements are issued by outage-probability constrained formulations from the prospective of probabilistic robust optimization, while the users' SINR requirements are guaranteed with all the CSI uncertainties from the prospective of deterministic robust optimization.  
		
		\item To strike a trade-off between system performance and coordination overhead, five corresponding low-complexity algorithms are proposed to minimize transmit power consumption for the three schemes in terms of probabilistic and deterministic manners. The Full-CI scheme based algorithms consume the least power consumption by utilizing both inter-cell and intra-cell interference; the Stat-CI based algorithm requires the smallest coordination overhead; and the Partial-CI scheme based algorithms make proper trade-off between power consumption and coordination overhead. The performance of the proposed algorithms are benchmarked by the existing CBF and CoMP. The complexities and coordination overhead of the algorithms are analytically demonstrated.

	\end{enumerate}

	\textit{Notations}:
	Matrices and vectors are represented by boldface capital and lower case letters, respectively. $\vert\cdot \vert$ denotes the absolute value of a complex scalar. $\vert\vert\cdot \vert\vert$ denotes the Euclidean vector norm. $\bm{A}^H$ $\bm{A}^T$ and Tr$(\bm{A})$ denote the Hermitian transpose, transpose and trace of matrix $\bm{A}$. Rank($\bm{A}$) denotes the rank of matrix $\bm{A}$. diag ($\bm{A}$) returns a diagonal matrix with diagonal elements from matrix $\bm{A}$ and diag ($\bm{a}$) stacks the elements of vector $\bm{a}$ into a diagonal matrix. $\bm{A}\succeq 0$ means $\bm{A}$ is a positive semi-definite matrix. Superscript $\Re$ and $\Im$ denote the real and imaginary parts, respectively. $\parallel \cdot \parallel_p$ means the p-norm of a vector or a matrix. $\bm{I}_n$ means an $n$-by-$n$ identity matrix. Operator vec$(\bm{A})$ stacks the element of matrix $\bm{A}$ into a vector.
	$\mathbb{C}^{N\times M}$ and $ \mathbb{H}^{N\times M}$  denote sets of all $N\times M$ matrices and Hermitian matrices with complex entries.

	\section{System Model and Constructive Interference  }
	In this section, system model is first introduced in II-A and then the concept of CI is briefly discussed in II-B.
	
	\subsection{System Model}
	We consider a multi-cell system at downlink transmission, where the coordinated BSs exchange CSI and/or data based on different schemes. Assume that each cell has one BS located in the center, and each BS is equipped with $M$ antennas for transmission. Without loss of generality, we assume there are $N_{BS}$ coordinated BSs and $K$ users in each cell. Each user is equipped with one antenna for simplicity.
	CSI is obtained by channel estimation in the training phase, based on channel reciprocity as in \cite{Wei2018Energy}.
	Define $\mathrm{U}_{ik}$ as the $k$-th user located in the $i$-th cell. Then its received signal can be expressed as
	
	\begin{small}
		\begin{equation}
		\begin{split}
		y_{ik}=\bm{h}_{iik}^T \sum_{n=1}^{K}\bm{w}_{in}s_{in}+\sum_{j\neq i}^{N_{BS}} \sum_{m=1}^{K} \bm{h}_{jik}^T \bm{w}_{jm}s_{jm} +n_{ik},
		\label{eq:Recevied signal 1}
		\end{split}
		\end{equation}
	\end{small}%
	where $h_{jik} \in \mathbb{C}^{M\times 1}$ represents the channel from the $j$-th BS to the user $\mathrm{U}_{ik}$. $w_{in}\in \mathbb{C}^{M\times 1}$ and $s_{in}$ denote the precoding and transmitted data at the $i$-th BS to the user $\mathrm{U}_{in}$.
	$n_{ik} \in \mathbb{C}$ denotes the white Gaussian additive Noise (WGAN) at the user $\mathrm{U}_{ik}$, following  $n_{ik} \sim \mathbb{CN}(0,\sigma_{n}^2), \forall i\in N_{BS}~\mathrm{and}~ \forall k\in K$.
	Conventionally by CBF, the SINR of the user $\mathrm{U}_{ik}$  is calculated as

	\begin{small}
		\begin{equation}
		\begin{split}
		\Gamma_{ik}^{\mathrm{CBF}}=\frac{  \left |  \bm{h}_{iik} \bm{w}_{ik}  \right | ^2    }{ \sum_{k'\neq k, k' \in i }^{ } \left |  \bm{h}_{iik} \bm{w}_{ik'}  \right | ^2  +     \sum_{j \neq i }^{N_{BS} }\sum_{m =1 }^{ K}  \left |  \bm{h}_{jik} \bm{w}_{jm}  \right | ^2         +\sigma_n^2},
		\label{eq:Conv SINR1}
		\end{split}
		\end{equation}
	\end{small}%
	where the first and second terms in the denominator represent the intra-cell (multi-user) and inter-cell interference, respectively. For the case of CoMP, the system essentially reduces to a network-level coordinated MIMO downlink, in which case the SINR is calculated as 
	
	\begin{small}
		\begin{equation}
		\begin{split}
		&\Gamma_{ik}^{\mathrm{CoMP}}=\frac{ \sum_{j=1}^{N_{BS}} \left |  \bm{h}_{jik} \bm{w}_{ik}  \right | ^2    }{ \sum_{j=1}^{N_{BS}}\sum_{k'\neq k, k' \in j }^{ }  \left |  \bm{h}_{jik'} \bm{w}_{ik'}  \right | ^2         +\sigma_n^2},
		\label{eq:Conv SINR2}
		\end{split}
		\end{equation}
	\end{small}%
	where the first term in the denominator represents the virtual intra-cell (multi-user) interference and needs to be mitigated. Evidently, CoMP serves multiple users through a broadcast channel.

	\subsection{Constructive Interference}
	
	\begin{figure}
		\centering
		\includegraphics[width=3.0 in]{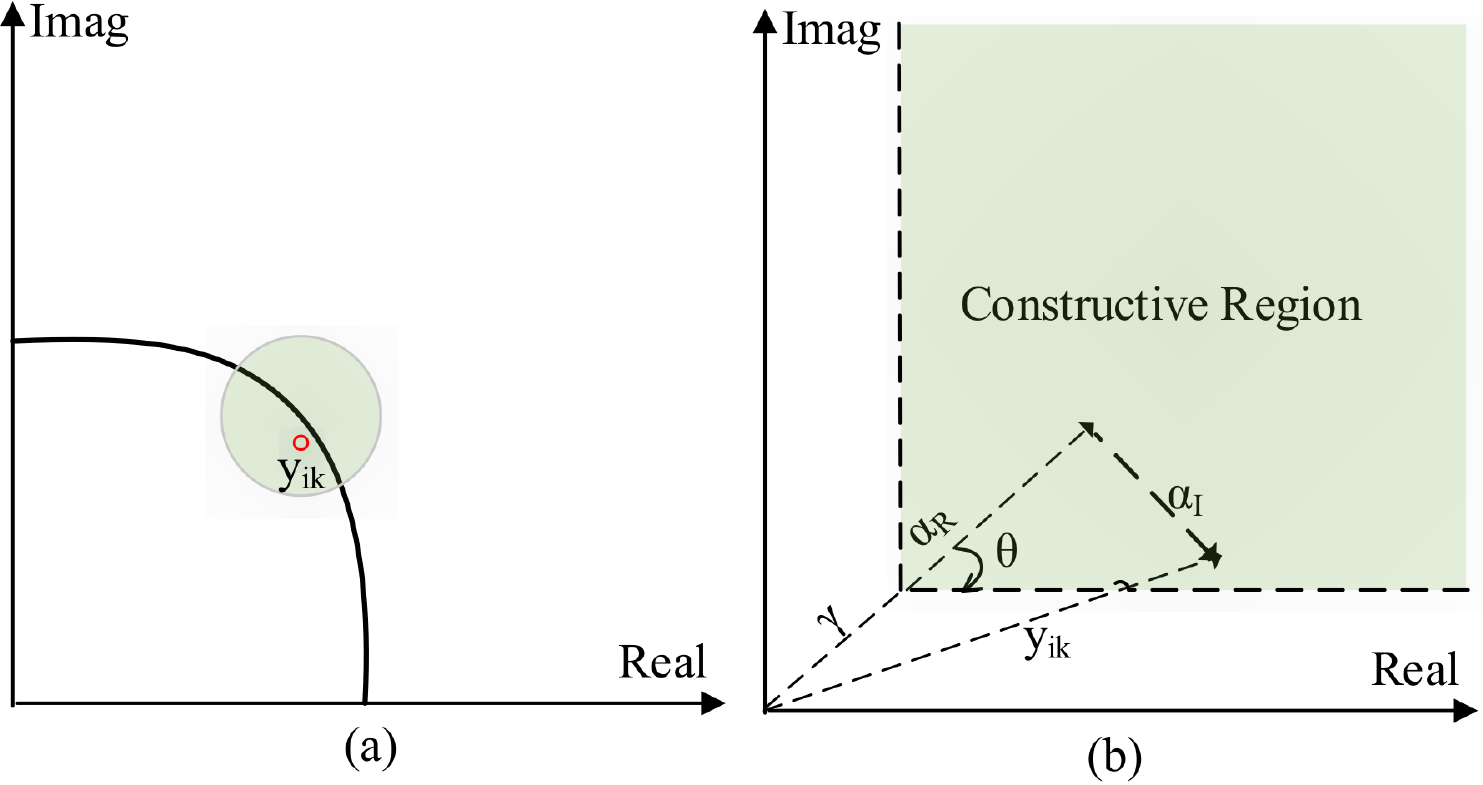}
		\caption{Optimization region for a) Conventional precoding. b) CI based precoding, where  QPSK modulation is used as an example and $\gamma=\sigma_n\sqrt{\Gamma_{ik}}$.}
		\label{fig:Region20181119}
	\end{figure}
	
	In CI, the interference is exploited to push the received signals away from the detection thresholds of the signal constellation. 
	The increased distance to the detection threshold can effectively improve the receiving performance. 
	The CI concept has been thoroughly discussed in the recent literature, and to avoid extensive repetition we refer the readers to \cite{Christos2011Correlation}. 
	Since the transmitted symbol can be written as $s_{ik}=d_{ik}e^{j(\phi_{ik})}$, for the purpose of the CI analysis it is convenient to express the symbol  by another symbol such that  $s_{in}=s_{ik}e^{j(\phi_{in}-\phi_{ik})}$. Hence, the received signal in (\ref{eq:Recevied signal 1}) can be re-written as 
	
	\begin{small}
		\begin{equation}
		\begin{split}
		y_{ik}=\bm{h}_{iik}^T \sum_{n=1}^{K} \bm{w}_{in}e^{(\phi_{in}-\phi_{ik})}s_{ik}+\sum_{j\neq i}^{N_{bs}}  \sum_{m=1}^{K} \bm{h}_{jik}^T \bm{w}_{jm} e^{(\phi_{jm}-\phi_{ik}})s_{ik} +n_{ik}.
		\label{eq:New Recevied signal 2}
		\end{split}
		\end{equation}
	\end{small}%

	Define $\tilde{\bm{h}_{jik}} =\bm{h}_{jik}e^{j(\phi_{11}-\phi_{ik})}$, and $\bm{w}_{j}=  \sum_{m=1 }^{K }\bm{w}_{jm}e^{j(\phi_{jm}-\phi_{11})}$, $\forall j\in N_{BS}$. Note that the  phase $\phi_{11}$ of the reference symbol $s_{11}$ can be arbitrary. Now,  the received signal at the user $\mathrm{U}_{ik}$ can be equivalently expressed as

	\begin{small}
		\begin{equation}
		\begin{split}
		y_{ik}= \sum_{j=1}^{N_{BS}} \tilde{\bm{h}_{jik}^T } \bm{w}_{j}+n_{ik}.
		\label{eq:New Recevied signal 3}
		\end{split}
		\end{equation}
	\end{small}%
	
	The reformulation in \eqref{eq:New Recevied signal 3} indicates that the original broadcast channel reduces to a virtual multicast channel with common messages to all users \cite{Timotheou2016Exploiting}. Hence,  by using the geometrical  interpretation in Fig. 1, the requirement of generating CI can be given as 
	
	\begin{small}
		\begin{equation}
		\begin{split}
		|\Im \big( { \sum_{j=1}^{N_{BS}}    \tilde{\bm{h}_{jik}^T } \bm{w}_{j}\big)  }| \leq  \big( { \Re(\sum_{j=1}^{N_{BS}}    \tilde{\bm{h}_{jik}^T }\bm{w}_{j} })-\sigma_n\sqrt{\Gamma_{ik}}\big)  \mathrm{tan}\theta, \forall~ \mathrm{U}_{ik},
		\label{eq:requirement of CI}
		\end{split}
		\end{equation}
	\end{small}%
	where the user's SINR requirement is also embedded. It indicates that generating CI and satisfying the user $\mathrm{U}_{ik}$'s SINR requirement are both guaranteed.

	\section{Fully-Coordinated CI Precoding with Imperfect CSI}
	
	In Section III, we  investigate the power efficient multi-cell CI design with imperfect CSI, where the channel from the $j$-th BS to the user U$_{ik}$ is given as $\tilde{\bm{h}_{jik}}=\hat{\bm{h}_{jik}}+\bm{e}_{ik}$, $\forall k \in K, j \in N_{BS}$. $\hat{\bm{h}_{jik}} \in \mathbb{C}^{M\times1}$ denotes the estimated channel with estimation error $\bm{e}_{ik} \in \mathbb{C}^{M\times1}$.
	The element of channel estimation error vector follows normal distribution such that $ [\bm{e}_{ik}]_m \sim \mathbb{CN}\{0,\sigma_{ik}^2\}$, $\forall m \in M$ \cite{Boshkovsha2015Practical}.
	To fully exploit the power efficient design with the uncertainties (CSI error), we handle the optimization problem in probabilistic and deterministic manners, respectively.


	\subsection{Probabilistic Robust Optimization}

	\subsubsection{Problem Formulation}
	Define precoding vector $\bm{w}_j \in \mathbb{C}^{M\times1}$, $\forall j \in N_{BS}$, as the precoding vector at the $j$-th BS. The optimization problem can be given as 
	
	\begin{small}
		\begin{equation}
		\begin{split}
		&P1~ (\mathrm{Full-CI-Prob}): \operatorname*{argmin}\limits_{\bm{w_j}, \forall j \in N_{BS}} \sum_{j=1}^{N_{BS}} \left \| w_j  \right \|    ^2      ,\\
		& \mathrm{s.t}~(C1):\left \| w_j  \right \|    ^2 \leq p_{max}, \forall j \in N_{BS},~(C2): \mathrm{Pr}\{\ \Gamma_{ik} \geq \overline{ \Gamma_{ik} } |  \bm{e}_{ik}   \} \geq \eta_{ik}, \forall i\in N_{BS}, k \in K,
		\label{eq:probabilistic_P1}
		\end{split}
		\end{equation}
	\end{small}%
	where $\overline{ \Gamma_{ik} } $ denotes the SINR requirement for the user $\mathrm{U}_{ik}$. $\eta_{ik}$ denotes the probabilistic threshold for the user $\mathrm{U}_{ik}$. Evidently, $(C1)$ imposes individual transmission power constraint at each BS. $(C2)$ guarantees that the SINR constraints at the user with probabilities $\eta_{ik}$ greater than $\overline{ \Gamma_{ik} }$.

	\subsubsection{Solution to the Problem}
	To solve the problem, at first we need to handle the probabilistic constraint $(C2)$. Under the provision of CI, the constraint $(C2)$ equals to
	
	\begin{small}
		\begin{equation}
		\begin{split}
		& (C2): \mathrm{Pr}\{\ \Gamma_{ik} \geq \overline{ \Gamma_{ik} } |  \bm{e}_{ik}   \} \geq \eta_{ik} \overset{(\ref{eq:requirement of CI})}{\Rightarrow} \\
		&   \mathrm{Pr}\{  | \sum_{j=1}^{N_{BS}}   (\tilde{\bm{h}_{jik}^{\Re}})^T \bm{w_j}^{\Im} +(\tilde{\bm{h}_{jik}^{\Im}}   )^T \bm{w_j}^{\Re} | \leq     \big(  \sum_{j=1}^{N_{BS}} \big( (\tilde{\bm{h}_{jik}^{\Re}})^T \bm{w}_j^{\Re} -(\tilde{\bm{h}_{jik}^{\Im} }  )^T \bm{w}_j^{\Im}  \big) -\sigma_n \sqrt{ \overline {\Gamma_{ik}} } \big)       \mathrm{tan}\theta | \bm{e}_{ik} \} \geq \eta_{ik},
		\label{eq:equivalent SINR}
		\end{split}
		\end{equation}
	\end{small}%
	
	Removing the absolute operator in \eqref{eq:equivalent SINR}, (C2) is equivalent to the two following inequalities
	
	\begin{small}
		\begin{equation}
		\left\{  
		\begin{array}{lr}  
		\mathrm{Pr}\{  \sum_{j=1}^{N_{BS}} \big(  (\tilde{\bm{h}_{jik}^{\Re}})^T \bm{w}_j^{\Im} +(\tilde{\bm{h}_{jik}^{\Im}}   )^T \bm{w}_j^{\Re} \big)\leq  
		\big(  \sum_{j=1}^{N_{BS}}\big( (\tilde{\bm{h}_{jik}^{\Re}})^T \bm{w}_j^{\Re} -(\tilde{\bm{h}_{jik}^{\Im} }  )^T \bm{w}_j^{\Im}  \big) -\sigma_n \sqrt{ \overline {\Gamma_{ik}} } \big)       \mathrm{tan}\theta | \bm{e}_{ik} \} \geq \eta_{ik},&\\
		\mathrm{Pr}\{ - \sum_{j=1}^{N_{BS}} \big(  (\tilde{\bm{h}_{jik}^{\Re}})^T \bm{w}_j^{\Im} + (\tilde{\bm{h}_{jik}^{\Im}}   )^T \bm{w}_j^{\Re} \big)\leq   
		\big(  \sum_{j=1}^{N_{BS}} \big( (\tilde{\bm{h}_{jik}^{\Re}})^T \bm{w_j}^{\Re} -(\tilde{\bm{h}_{jik}^{\Im} }  )^T \bm{w_j}^{\Im} \big)  -\sigma_n \sqrt{ \overline {\Gamma_{ik}} } \big)       \mathrm{tan}\theta | \bm{e}_{ik} \} \geq \eta_{ik}.
		\end{array}  
		\right.  
		\label{eq:equivalent SINR 2}
		\end{equation}
	\end{small}%
	
	We now focus our attention on the first inequality, which can be expanded into
	

	\begin{small}
		\begin{equation}  
		\begin{split}
		& \mathrm{Pr}\{  \sum_{j=1}^{N_{BS}} \big( [ \hat{\bm{h}_{jik}^{\Im}}+\bm{e}_{ik}^{\Im}  - \hat{\bm{h}_{jik}^{\Re}}\mathrm{tan}\theta- \bm{e}_{ik}^{\Re}\mathrm{tan}\theta; \hat{\bm{h}_{jik}^{\Re}}+\bm{e}_{ik}^{\Re} \\ 
		&~~~~~~~~~~  + \hat{\bm{h}_{jik}^{\Im}}\mathrm{tan}\theta+ \bm{e}_{ik}^{\Im}\mathrm{tan}\theta]^T [\bm{w_j}^{\Re}; \bm{w_j}^{\Im}]\big)    \leq -\sigma_n\sqrt{ \overline{  \Gamma_{ik}}}\mathrm{tan}\theta                       \} \geq \eta_{ik}.
		\label{eq:CI probability 2}
		\end{split}
		\end{equation} 
	\end{small}%
	
	For simplicity, we write \eqref{eq:CI probability 2} as $\mathrm{Pr}\{ \sum_{j=1}^{N_{BS}} \big( \bm{a}_{jik}^T[\bm{w}_j^{\Re};\bm{w}_j^{\Im}] \big) \leq -\sigma_n\sqrt{ \overline{  \Gamma_{ik}}}\mathrm{tan}\theta| \bm{e}_{ik}\}  \geq \eta_{ik}$, where $ \bm{a}_{jik}=[ \hat{\bm{h}_{jik}^{\Im}}+\bm{e}_{ik}^{\Im}  - \hat{\bm{h}_{jik}^{\Re}}\mathrm{tan}\theta- \bm{e}_{ik}^{\Re}\mathrm{tan}\theta; \hat{\bm{h}_{jik}^{\Re}}+\bm{e}_{ik}^{\Re}  + \hat{\bm{h}_{jik}^{\Im}}\mathrm{tan}\theta+ \bm{e}_{ik}^{\Im}\mathrm{tan}\theta]$. 
	It is easy to obtain that the $M$-dimensional normal distributed vector $\bm{a}_{jik}$'s expectation is $ \overline{\bm{a}_{jik}} =[  \hat{\bm{h}_{jik}^{\Im}}  - \hat{\bm{h}_{jik}^{\Re}}\mathrm{tan}\theta; \hat{\bm{h}_{jik}^{\Re}} + \hat{\bm{h}_{jik} ^{\Im}}\mathrm{tan}\theta]$ with covariance matrix $\bm{\Theta}_{jik}=\mathrm{diag} ( \underbrace{ (1+\mathrm{tan}^2\theta)\sigma_{ik}^2,...,(1+\mathrm{tan}\theta)^2\sigma_{ik}^2}_{2M}) $. 
	We now stack the vectors $\bm{w}_j$ and $\bm{a}_{jik}, \forall j \in N_{BS},$ into two long vectors $\bm{w}$ and $\bm{a}_{ik}, \forall j \in N_{BS},$ such that $\bm{w}=\mathrm{vec}( \bm{w}_1^{\Re};\bm{w}_1^{\Im};\bm{w}_2^{\Re};\bm{w}_2^{\Im};...;\bm{w}^{\Re}_{N_{BS}};\bm{w}^{\Im}_{N_{BS}})$ and $\bm{a}_{ik}=\mathrm{vec}(\bm{a}_{1ik};\bm{a}_{1ik};...;\bm{a}_{N_{BS}ik})$. It is observed that the long vector $\bm{a}_{ik}$'s expectation is calculated as $ \overline{\bm{a}_{ik}}=\mathrm{vec}(\overline{\bm{a}_{1ik}};\overline{\bm{a}_{1ik}};...;\overline{\bm{a}_{N_{BS}ik}})$ with  covariance matrix $\bm{\Theta}_{ik}=\mathrm{diag} ( \underbrace{ (1+\mathrm{tan}^2\theta)\sigma_{ik}^2,...,(1+\mathrm{tan}^2\theta)\sigma_{ik}^2}_{2M \times N_{BS}}) $. Now the constraint in \eqref{eq:CI probability 2} can be derived as

	

	\begin{small}
		\begin{equation}
		\begin{split}
		& \mathrm{Pr}\{\  \bm{a}_{ik}^T[ \bm{w}_1^{\Re};\bm{w}_1^{\Im};...;\bm{w}^{\Re}_{N_{BS}};\bm{w}^{\Im}_{N_{BS}}] \leq -\sigma_n\sqrt{ \overline{  \Gamma_{ik}}}\mathrm{tan}\theta |  \bm{e}_{ik}   \} \geq \eta_{ik}  \Rightarrow\\
		&\mathrm{Pr}\{ \frac{ \bm{a}_{ik}^T[\bm{w}_1^{\Re};\bm{w}_1^{\Im};...;\bm{w}^{\Re}_{N_{BS}};\bm{w}^{\Im}_{N_{BS}}]- \overline{\bm{a}_{ik}}^T[\bm{w}_1^{\Re};\bm{w}_1^{\Im};...;\bm{w}^{\Re}_{N_{BS}};\bm{w}^{\Im}_{N_{BS}}]}{ || \bm{\Theta}^{\frac{1}{2}}_{ik} [\bm{w}_1^{\Re};\bm{w}_1^{\Im};...;\bm{w}^{\Re}_{N_{BS}};\bm{w}^{\Im}_{N_{BS}}]||_2} \\
		& ~~~~~~~  \leq \frac{ -\sigma_n\sqrt{ \overline{  \Gamma_{ik}}}\mathrm{tan}\theta-\overline{\bm{a}_{ik}}^T [\bm{w}_1^{\Re};\bm{w}_1^{\Im};...;\bm{w}^{\Re}_{N_{BS}};\bm{w}^{\Im}_{N_{BS}}]}{|| \bm{\Theta}^{\frac{1}{2}}_{ik} [\bm{w}_1^{\Re};\bm{w}_1^{\Im};...;\bm{w}^{\Re}_{N_{BS}};\bm{w}^{\Im}_{N_{BS}}] ||_2}   |  \bm{e}_{ik}   \} \geq \eta_{ik} \\
		& \Rightarrow \Phi(  \frac{ -\sigma_n \sqrt{ \overline{  \Gamma_{ik}}}\mathrm{tan}\theta- \overline{\bm{a}_{ik}}^T[\bm{w}_1^{\Re};\bm{w}_1^{\Im};...;\bm{w}^{\Re}_{N_{BS}};\bm{w}^{\Im}_{N_{BS}}]}{|| \bm{\Theta}^{\frac{1}{2}}_{ik} [\bm{w}_1^{\Re};\bm{w}_1^{\Im};...;\bm{w}^{\Re}_{N_{BS}};\bm{w}^{\Im}_{N_{BS}}] ||_2}   ) \geq \eta_{ik},
		\label{eq:CI probability 3}
		\end{split}
		\end{equation}
	\end{small}%
	where $\Phi(x)=\frac{2}{\sqrt{\pi}} \int_{0}^{x}e^{-t^2} \mathrm{d}t$ denotes the cumulative probability function (cdf) of a standard normal distributed variable. Defining $\Phi^{-1}(^.)$ as the inverse function of $\Phi(^.)$, (\ref{eq:CI probability 3}) can be finally derived into a quadratic constraint such that
	
	\begin{small}
		\begin{equation}
		\begin{split}
		\overline{\bm{a}_{ik}}^T [\bm{w}_1^{\Re};\bm{w}_1^{\Im};...;\bm{w}^{\Re}_{N_{BS}};\bm{w}^{\Im}_{N_{BS}}]+ \Phi^{-1}(\eta_{ik} )||  \bm{\Theta}^{\frac{1}{2}}_{ik,1} [\bm{w}_1^{\Re};\bm{w}_1^{\Im};...;\bm{w}^{\Re}_{N_{BS}};\bm{w}^{\Im}_{N_{BS}}] ||_2 \leq  -\sigma_n\sqrt{ \overline{  \Gamma_{ik}}}\mathrm{tan}\theta.
		\label{eq:CI probability 4}
		\end{split}
		\end{equation}
	\end{small}%
	
	Similarly, the second inequality in (\ref{eq:equivalent SINR 2}) can be given as

	\begin{small}
		\begin{equation}
		\begin{split}
		\overline{\bm{b}_{ik}}^T  [\bm{w}_1^{\Re};\bm{w}_1^{\Im};...;\bm{w}^{\Re}_{N_{BS}};\bm{w}^{\Im}_{N_{BS}}]+ \Phi^{-1}(\eta_{ik} )|| \bm{\Theta}^{\frac{1}{2}}_{ik} [\bm{w}_1^{\Re};\bm{w}_1^{\Im};...;\bm{w}^{\Re}_{N_{BS}};\bm{w}^{\Im}_{N_{BS}}] ||_2 
		\leq  -\sigma_n\sqrt{ \overline{  \Gamma_{ik}}}\mathrm{tan}\theta,
		\label{eq:CI probability 5}
		\end{split}
		\end{equation}
	\end{small}%
	where $\overline{\bm{b}_{ik}} =\mathrm{vec} (\overline{\bm{b}_{1ik}}; \overline{\bm{b}_{2ik}};...;\overline{\bm{b}_{N_{BS}ik}})$ and $\overline{\bm{b}_{jik}}=[- \hat{ \bm{h}_{jik}^{\Im}}-\hat{ \bm{h}_{jik}^{\Re}} \mathrm{tan}\theta;-\hat{ \bm{h}_{jik}^{\Re}}+\hat{ \bm{h}_{jik}^{\Im}} \mathrm{tan}\theta]$ with covariance matrix calculated as $\bm{\Theta}_{ik}=\mathrm{diag} ( \underbrace{(1+\mathrm{tan}\theta)^2\sigma_{ik}^2,...,(1+\mathrm{tan}\theta)^2\sigma_{ik}^2}_{2M\times N_{BS}})  $.  
	Now,  constraint $(C2)$ has been transformed into the two inequalities in \eqref{eq:CI probability 4} and \eqref{eq:CI probability 5}.
	According to Schur Complements that $ ||\bm{A}\bm{x}+\bm{b}||_2\leq \bm{e}^T\bm{x}+d$  is equivalent to 
	$\left[
	\begin{smallmatrix}
	(\bm{e}^T\bm{x}+d)\bm{I}& \bm{A}\bm{x}+\bm{b}\\
	(\bm{A}\bm{x}+\bm{b})^T& \bm{e}^T\bm{x}+d\\
	\end{smallmatrix}  
	\right]\succeq  \bm{0}$   \cite{Boyd2004Convex}, the above two constraints can be further transformed into two linear matrix inequalities (LMI) as

	\begin{equation}  
	\left\{  
	\begin{array}{lr}  
	\left[
	\begin{smallmatrix}
	\frac{ \big(- \overline{\bm{a}_{ik}}^T [\bm{w}_1^{\Re};\bm{w}_1^{\Im};...;\bm{w}^{\Re}_{N_{BS}};\bm{w}^{\Im}_{N_{BS}}]-\sigma_n\sqrt{ \overline{  \Gamma_{ik}}}\mathrm{tan}\theta\big) \bm{I}}{\Phi^{-1}(\eta_{ik}) }&   \bm{\Theta}^{\frac{1}{2}}_{ik} [\bm{w}_1^{\Re};\bm{w}_1^{\Im};...;\bm{w}^{\Re}_{N_{BS}};\bm{w}^{\Im}_{N_{BS}}]\\
	(\bm{\Theta}^{\frac{1}{2}}_{ik} [\bm{w}_1^{\Re};\bm{w}_1^{\Im};...;\bm{w}^{\Re}_{N_{BS}};\bm{w}^{\Im}_{N_{BS}}]) ^T& \frac{-\overline{\bm{a}_{ik}}^T [\bm{w}_1^{\Re};\bm{w}_1^{\Im};...;\bm{w}^{\Re}_{N_{BS}};\bm{w}^{\Im}_{N_{BS}}]-\sigma_n\sqrt{ \overline{  \Gamma_{ik}}}\mathrm{tan}\theta}{\Phi^{-1}(\eta_{ik}) } \\
	\end{smallmatrix}
	\right]  \succeq \bm{0}, &  \\  
	\\
	\left[
	\begin{smallmatrix}
	\frac{ \big(-\overline{\bm{b}_{ik}}^T [\bm{w}_1^{\Re};\bm{w}_1^{\Im};...;\bm{w}^{\Re}_{N_{BS}};\bm{w}^{\Im}_{N_{BS}}]-\sigma_n\sqrt{ \overline{  \Gamma_{ik}}}\mathrm{tan}\theta\big) \bm{I}}{\Phi^{-1}(\eta_{ik}) }&  \bm{\Theta}^{\frac{1}{2}}_{ik} [\bm{w}_1^{\Re};\bm{w}_1^{\Im};...;\bm{w}^{\Re}_{N_{BS}};\bm{w}^{\Im}_{N_{BS}}] \\
	(\bm{\Theta}^{\frac{1}{2}}_{ik} [\bm{w}_1^{\Re};\bm{w}_1^{\Im};...;\bm{w}^{\Re}_{N_{BS}};\bm{w}^{\Im}_{N_{BS}}])^T& \frac{-\overline{\bm{b}_{ik}}^T [\bm{w}_1^{\Re};\bm{w}_1^{\Im};...;\bm{w}^{\Re}_{N_{BS}};\bm{w}^{\Im}_{N_{BS}}]-\sigma_n\sqrt{ \overline{  \Gamma_{ik}}}\mathrm{tan}\theta}{\Phi^{-1}(\eta_{ik}) } \\
	\end{smallmatrix}
	\right] \succeq \bm{0}
	\end{array}  
	\right.  
	\label{eq:Schur1}
	\end{equation}

	Now we define $\bm{W}_j=\bm{w}_j^H\bm{w}_j, \forall j \in N_{BS}$. P1 can be equivalently written as


	\begin{small}
		\begin{equation}
		\begin{split}
		& P2~ (\mathrm{Full-CI-Prob}): \operatorname*{argmin}\limits_{\bm{W}_j, \in N_{BS}} \sum_{j=1}^{N_{BS}} \mathrm{Tr} (\bm{W}_j) ,\\
		&\mathrm{s.t~}(C1):  \mathrm{Tr}(\bm{W}_j) \leq p_{max}, \forall j \in N_{BS},~ (C2):  (\ref{eq:Schur1}),~(C3):  \left[
		\begin{matrix}
		\bm{W}_j& \bm{w}_j\\
		\bm{w}_j^T& 1
		\end{matrix}
		\right] \succeq 0, \forall j \in N_{BS},
		\label{eq:CI probability P2}
		\end{split}
		\end{equation}
	\end{small}%
	which is a convex semi-definite programming (SDP) problem and can be readily solved by CVX solver.

	\subsection{Deterministic Robust Optimization}
	
	In the previous subsection, we have solved the problem in a probabilistic manner, where the users' QoS requirements are issued by the chance constrained formulations. In this section, we handle the CSI uncertainties in a deterministic manner, where the users' QoS requirements are satisfied all the time with the infinite CSI uncertainties.
	Define $\bm{\Delta}$ as the channel estimation uncertainties set, which contains all the possible CSI uncertainties and specifies an ellipsoidal uncertainty region for the estimated CSI \cite{Boshkovsha2015Practical}. 
	\subsubsection{Problem Formulation}
	To process the power minimization problem in terms of deterministic robust optimization, the formulation is given as 
	
	\begin{small}
		\begin{equation}
		\begin{split}
		& P3~ (\mathrm{Full-CI-Det}): \operatorname*{argmin}\limits_{\bm{w}_j,j\in N_{BS}}  \sum_{j=1}^{N_{BS}} \left \| \bm{w}_j  \right \|    ^2 ,\\
		&\mathrm{s.t~}(C4): \left \| \bm{w}_j  \right \|    ^2 \leq p_{max}, \forall j\in N_{BS},~(C5):  \operatorname*{min}\limits_{\bm{e_{ik}\in \bm{\Delta}}} \Gamma_{ik} \geq \overline{ \Gamma_{ik} }, \forall i\in N_{BS}, k\in K,  
		\label{eq:CI det p1}
		\end{split}
		\end{equation}
	\end{small}%
	where $(C5)$ indicates the deterministic SINR requirement for the users, such that the users' worst-case SINR as per the CSI error distribution obey the respective thresholds $\overline{\Gamma_{ik}}$.

	\subsubsection{Optimization Solution}
	
	In line with the analysis in previous section, constraint $(C5)$ is equivalent to the following two inequalities

	\begin{small}
		\begin{equation}  
		\left\{  
		\begin{array}{lr}  
		\operatorname*{min}\limits_{\bm{e_{ik}\in \Delta}} \sum_{j=1}^{N_{BS}}  \big( (\tilde{\bm{h}_{jik}^{\Re}})^T \bm{w_j}^{\Im} +(\tilde{\bm{h}_{jik}^{\Im}}   )^T \bm{w_j}^{\Re} \big)\leq 
		\big(  \sum_{j=1}^{N_{BS}} \big( (\tilde{\bm{h}_{jik}^{\Re}})^T \bm{w_j}^{\Re} -(\tilde{\bm{h}_{ijk}^{\Im}}   )^T \bm{w_j}^{\Im}\big)   -\sigma_n \sqrt{ \overline {\Gamma_{ik}} } \big)       \mathrm{tan}\theta,\\
		\operatorname*{min}\limits_{\bm{e_{ik}\in \Delta}} - \sum_{j=1}^{N_{BS}}  \big( (\tilde{\bm{h}_{jik}^{\Re}})^T \bm{w_j}^{\Im} +(\tilde{\bm{h}_{jik}^{\Im}}   )^T \bm{w_j}^{\Re} \big) \leq   
		\big(  \sum_{j=1}^{N_{BS}} \big( (\tilde{\bm{h}_{jik}^{\Re}})^T \bm{w_j}^{\Re} -(\tilde{\bm{h}_{jik}^{\Im}}   )^T \bm{w_j}^{\Im}  \big) -\sigma_n \sqrt{ \overline {\Gamma_{ik}} } \big)       \mathrm{tan}\theta.
		\end{array}  
		\right.  
		\label{eq:CI det SINR}
		\end{equation}  
	\end{small}%
	
	We focus our attention on the first inequality of (\ref{eq:CI det SINR}), which can be further written as 
	
	\begin{small}
		\begin{equation}  
		\begin{split}
		\operatorname*{min}\limits_{\bm{e_{ik}\in \Delta}}  \sum_{j=1}^{N_{BS}} \big( [  \bm{e}_{ik}^{\Im}-\bm{e}_{ik}^{\Re}\mathrm{tan\theta}; \bm{e}_{ik}^{\Re}+\bm{e}_{ik}^{\Im}\mathrm{tan\theta}    ]^T [ \bm{w_j}^{\Re}, \bm{w_j}^{\Im}]+\rho_{ik}\big) \leq 0,
		\label{eq:CI det SINR 2}
		\end{split}
		\end{equation}  
	\end{small}%
	where $\rho_{ik}=  \sum_{j=1}^{N_{BS}} \big( (\hat{\bm{h}_{jik}^{\Im}})^T \bm{w}_j^{\Re}-(\hat{\bm{h}_{jik}^{\Re}})^T  \bm{w}_j^{\Re} \mathrm{tan\theta}+(\hat{\bm{h}_{jik}^{\Re}} )^T \bm{w}_j^{\Im}+(\hat{\bm{h}_{jik}^{\Im}})^T  \bm{w}_j^{\Im} \mathrm{tan\theta}\big)+ \sigma_n\sqrt{\overline{\Gamma_{ik}}}\mathrm{tan\theta}$. For simplicity, define vector $\bm{c}_{ik}=\mathrm{vec}( \bm{c}_{1ik}; \bm{c}_{2ik};...;\bm{c}_{N_{BS}ik})$, where $\bm{c}_{jik}=[  \bm{e}_{ik}^{\Im}-\bm{e}_{ik}^{\Re}\mathrm{tan\theta}; \bm{e}_{ik}^{\Re}+\bm{e}_{ik}^{\Im}\mathrm{tan\theta}    ]$. Then Eq. \eqref{eq:CI det SINR 2} becomes
	
	\begin{small}
		\begin{equation}  
		\begin{split}
		&\operatorname*{min}\limits_{\bm{e_{ik}\in \Delta}} \bm{c}_{ik}^T [\bm{w}_1^{\Re};\bm{w}_1^{\Im};...;\bm{w}^{\Re}_{N_{BS}};\bm{w}^{\Im}_{N_{BS}}]+ \rho_{ik} \leq 0.
		\label{eq:CI det SINR 4}
		\end{split}
		\end{equation}  
	\end{small}%

	To handle the infinite CSI uncertainties in \eqref{eq:CI det SINR 4}, we transform it into a LMI using the following Lemma 1 (S-procedure):
	
	\textbf{\textit{Lemma 1}} (S-Procedure \cite{Boyd2004Convex}): Let a function $ f_m({\bm{x}})$, $m \in \{1,2\}$, be defined as
	
	\begin{small}
		\begin{equation}
		\begin{split}
		f_m({\bm{x}})=\bm{x}^H\bm{A}_m\bm{x}+2\Re \{   \bm{b}_m^H\bm{x}  \}+c_m
		\label{eq:determinstic equivalent IR3}
		\end{split}
		\end{equation}
	\end{small}%
	where $\bm{A}_m \in \mathbb{H}^{N\times N}$, $\bm{b}_m \in \mathbb{C}^{N\times1}$ and $c_m \in \mathbb{R}$. The implication $ f_1({\bm{x}}) \Rightarrow  f_2({\bm{x}})$ holds if and only if there exists a $\lambda$ such that
	
	\begin{small}
		\begin{equation}
		\begin{split}
		\lambda \left[
		\begin{matrix}
		\bm{A}_1,& \bm{b}_1\\
		\bm{b}_1^H,& c_1
		\end{matrix}
		\right]-
		\left[
		\begin{matrix}
		\bm{A}_2,& \bm{b}_2\\
		\bm{b}_2^H,& c_2
		\end{matrix}
		\right] \succeq \bm{0}.
		\label{eq:determinstic equivalent IR5}
		\end{split}
		\end{equation}
	\end{small}%
	
	
	


	To utilize S-procedure, we need to construct a premise that guarantees \eqref{eq:CI det SINR 4} hold. By examining the structure of \eqref{eq:CI det SINR 4}, the premise can be evidently constructed as $(\sqrt{\bm{c}_{ik}}) ^T\bm{I}_{2N\times N_{BS}} \sqrt{\bm{c}_{ik}} \leq \xi^2$, and the  value of $\xi^2$ will be given by Lemma 2.  By S-procedure, to guarantee the implication in \eqref{eq:CI det SINR 6}  holds

	\begin{small}
		\begin{equation}
		\begin{split}
		&(\sqrt{\bm{c}_{ik}}) ^T\bm{I}_{2M\times N_{BS}} \sqrt{\bm{c}_{ik}}  - \xi^2 \leq 0 \Rightarrow (\sqrt{\bm{c}_{ik}}) ^T \mathrm{diag} ( [\bm{w}_1^{\Re};\bm{w}_1^{\Im};...;\bm{w}^{\Re}_{N_{BS}};\bm{w}^{\Im}_{N_{BS}}])\sqrt{\bm{c}_{ik}} + \rho_{ik} \leq 0,
		\label{eq:CI det SINR 6}
		\end{split}
		\end{equation}
	\end{small}
	the following LMI constraint in \eqref{eq:CI det SINR 7} should hold with $\lambda_{ik}\geq 0$

	\begin{small}
		\begin{equation}
		\begin{split}
		&\lambda_{ik} \left[
		\begin{matrix}
		\bm{I}_{2M\times N_{BS}},& \bm{0}\\
		\bm{0},& -\xi^2
		\end{matrix}
		\right]-
		\left[
		\begin{matrix}
		\mathrm{diag}( [\bm{w}_1^{\Re};\bm{w}_1^{\Im};...;\bm{w}^{\Re}_{N_{BS}};\bm{w}^{\Im}_{N_{BS}}]),& \bm{0}\\
		\bm{0},&  \rho_{ik}\\
		\end{matrix}
		\right] \succeq \bm{0}, 
		\label{eq:CI det SINR 7}
		\end{split}
		\end{equation}
	\end{small}%
	by which the first inequality of \eqref{eq:CI det SINR} containing infinite possibilities is transformed into a deterministic LMI. However, variable $\xi^2$ is introduced to bound the term $(\sqrt{\bm{c}_{ik}}) ^T\bm{I}_{2N\times N_{BS}} \sqrt{\bm{c}_{ik}}$, which needs to be connected with the known channel estimation error variance $\sigma_{ik}^2$. Hence, we now introduce Lemma 2 as follows.

	\textbf{\textit{Lemma 2}} (links $\xi^2$ to channel estimation error variance $\sigma_{ik}^2$): 
	Provided that the element of CSI error follows normal distribution such that $[\bm{e}_{ik}]_m \sim \mathcal{CN}(0,\sigma_{ik}^2)$, $\xi^2=\Phi^{-1}(\delta)\sqrt{  MN_{BS}(1+\mathrm{tan}^2\theta)      }\sigma_{ik}$ is obtained. $\delta$ physically represents the outage probability of $(\sqrt{\bm{c}_{ik}}) ^T\bm{I}_{2N\times N_{BS}} \sqrt{\bm{c}_{ik}}$ violating its upper bound $\xi^2$,  which can be set close to 1, i.e.,  $\delta=0.99$ \cite{Nasseri2016Chance}.  $\Phi^{-1}(\cdot)$ denotes the inverse function of the cdf of a standard normal distributed variable.

	Proof: Please see Appendix A. $\blacksquare$

	Now with the known value of $\xi^2$, the deterministic LMI constraint in \eqref{eq:CI det SINR 7} is solvable. Similarly, the second inequality in (\ref{eq:CI det SINR}) can be transformed as

	\begin{small}
		\begin{equation}
		\begin{split}
		&\omega_{ik} \left[
		\begin{matrix}
		\bm{I}_{2N\times N_{BS}},& \bm{0}\\
		\bm{0},& -\xi^2
		\end{matrix}
		\right]-\left[
		\begin{matrix}
		\mathrm{diag}( [\bm{w}_1^{\Re};\bm{w}_1^{\Im};...;\bm{w}^{\Re}_{N_{BS}};\bm{w}^{\Im}_{N_{BS}}]),& \bm{0}\\
		\bm{0},&  g_{ik}\\
		\end{matrix}
		\right] \succeq \bm{0}, \forall i \in {N_{BS}}, k \in {K}.
		\label{eq:CI det SINR 8}
		\end{split}
		\end{equation}
	\end{small}%
	where $g_{ik}=  \sum_{j=1}^{N_{BS}} \big( -(\hat{\bm{h}_{jik}^{\Im}})^T \bm{w}_j^{\Re}-(\hat{\bm{h}_{jik}^{\Re}})^T  \bm{w}_j^{\Re} \mathrm{tan\theta}-(\hat{\bm{h}_{jik}^{\Re}} )^T \bm{w}_j^{\Im}+(\hat{\bm{h}_{jik}^{\Im}})^T  \bm{w}_j^{\Im} \mathrm{tan\theta}\big)+ \sigma_n\sqrt{\overline{\Gamma_{ik}}}\mathrm{tan\theta}$.
	Now the constraint $(C5)$ containing infinite possibilities is transformed into two deterministic LMI inequalities in (\ref{eq:CI det SINR 7}) and (\ref{eq:CI det SINR 8}), respectively.
	Defining $\bm{W}=\bm{w}\bm{w}^H$, P3 can be equivalently written as

	\begin{small}
		\begin{equation}
		\begin{split}
		& P4~ (\mathrm{Full-CI-Det}): \operatorname*{argmin}\limits_{\bm{W}_j, \forall j \in N_{BS}} \sum_{j=1}^{N_{BS}}\mathrm{Tr}(\bm{W}_j),\\
		&\mathrm{s.t}~(C7):  \mathrm{Tr}(\bm{W}_j) \leq p_{max}, \forall j \in N_{BS},~(C8):  (\ref{eq:CI det SINR 7})~ \mathrm{and}~ (\ref{eq:CI det SINR 8}),\\
		&~~~~~(C9): \lambda_{ik} \geq 0,~(C10): \omega_{ik} \geq 0, \forall i \in N_{BS}, k \in K,~(C11):  \left[
		\begin{matrix}
		\bm{W}_j& \bm{w}_j\\
		\bm{w}_j^T& 1
		\end{matrix}
		\right] \succeq 0,\forall j \in N_{BS},
		\label{eq:CI det P2}
		\end{split}
		\end{equation}
	\end{small}%
	which is ready to solve as a standard SDP problem.

	\section{Partially-Coordinated CI Precoding with Imperfect CSI}
	
	In the previous section, both the intended transmitted data  and CSI need to be shared among the coordinated BSs to exploit the constructive inter-cell and intra-cell interference, which generally requires that the coordinated BSs are connected with high-capacity and delay-free links.
	In this section, to reduce the coordination overhead, we propose a scheme that the coordinated BSs only need to share CSI with others yet utilizing intra-cell interference as CI, which is referred as Partial-CI scheme.
	Since only intra-cell interference is designed as CI while inter-cell interference is considered as a harmful element, the received signal at user $\mathrm{U_{ik}}$ can be written as

	\begin{small}
		\begin{equation}
		\begin{split}
		y_{ik}=\bm{h}_{iik}^T \sum_{n=1}^{K}\bm{w}_{in}s_{in}+ \sum_{j \neq i}^{N_{BS}} \bm{h}_{jik}^T \sum_{m=1}^{K} \bm{w}_{jm}s_{jm} +n_{ik}.
		\label{eq:Received CI det}
		\end{split}
		\end{equation}
	\end{small}%

	Since we have $s_{in}=e^{j(\phi_{in}-\phi_{ik})} s_{ik}$ and $s_{jm}=e^{j(\phi_{jm}-\phi_{jk})} s_{jk}$, (\ref{eq:Received CI det}) can be expressed as
	
	\begin{small}
		\begin{equation}
		\begin{split}
		&y_{ik}=\bm{h}_{iik} e^{j(\phi_{i1}-\phi_{ik})}\sum_{n=1}^{K}\bm{w}_{in}e^{j(\phi_{in}-\phi_{i1})}s_{ik} + \sum_{j\neq i}^{N_{BS}} \bm{h}_{jik} e^{j(\phi_{j1}-\phi_{jk})} \sum_{m=1}^{K} \bm{w}_{jm}e^{j(\phi_{jm}-\phi_{j1})} s_{jk} +n_{ik}.
		\label{eq:Received CI det 2}
		\end{split}
		\end{equation}
	\end{small}%
	
	Define $\tilde{\bm{h}_{iik}}=\bm{h}_{iik}e^{j(\phi_{i1}-\phi_{ik})}$, $\tilde{\bm{h}_{jik}}=\bm{h}_{jik}e^{j(\phi_{j1}-\phi_{jk})}$, $\bm{w}_i=\sum_{n=1}^{K}w_{in}e^{j(\phi_{in}-\phi_{i1})}$, and $\bm{w}_j=\sum_{m=1}^{K}w_{jm}e^{j(\phi_{jm}-\phi_{j1})}$. Eq. (\ref{eq:Received CI det 2}) can be written as

	\begin{small}
		\begin{equation}
		\begin{split}
		y_{ik}=(\tilde{\bm{h}_{iik}})^T \bm{w}_{i} s_{ik}+ \sum_{j \neq i}^{N_{BS}} (\tilde{ \bm{h}_{jik}})^T\bm{w}_{j} s_{jk}+n_{ik}.
		\label{eq:Received CI det 3}
		\end{split}
		\end{equation}
	\end{small}%

	Hence, the SINR of the Partial-CI scheme can be written as 
	
	\begin{small}
		\begin{equation}
		\begin{split}
		\Gamma_{ik}^{\mathrm{Partial-CI}}=\frac{  \left |  \tilde{\bm{h}_{iik}^T} \bm{w}_{i}  \right | ^2    }{ \sum_{j\ne i}^{N_{BS}}  \left | \tilde{ \bm{h}_{jik}^T} \bm{w}_{j}  \right | ^2         +\sigma_n^2}.
		\label{eq:Received CI det 4}
		\end{split}
		\end{equation}
	\end{small}%
	
	To minimize power consumption in the scenario, we formulate and solve the problems in terms of probabilistic and deterministic manners, respectively.

	\subsection{Probabilistic Robust Optimization}
	In this subsection, we formulate and solve the problem in a probabilistic manner.
	
	\subsubsection{Problem Formulation}
	The problem is formulated as 
	
	\begin{small}
		\begin{equation}
		\begin{split}
		& P5~ (\mathrm{Partial-CI-Prob}):\operatorname*{argmin}\limits_{\bm{w}_j, \forall j \in N_{BS}}   \sum_{j=1}^{N_{BS}} ||  \bm{w}_j ||^2,\\
		&\mathrm{s.t}~(C11):||  \bm{w}_j  ||^2 \leq P_{max}, \forall j\in N_{BS},~(C12): \mathrm{Pr}\{\ \Gamma_{ik} \geq \overline{ \Gamma_{ik} } |  \bm{e}_{ik}   \} \geq \eta_{ik}.
		\label{eq:Partial CI det P5}
		\end{split}
		\end{equation}
	\end{small}%

	\subsubsection{Optimization Solution}
	Since inter-cell interference is treated as undesired element,  the probabilistic constraint (C12) equals to the following two inequalities

	\begin{small}
		\begin{equation}
		\left\{  
		\begin{array}{lr}  
		\mathrm{Pr}\{   (\tilde{\bm{h}_{iik}^{\Re}})^T \bm{w_i}^{\Im} +(\tilde{\bm{h}_{iik}^{\Im}}   )^T \bm{w_i}^{\Re} \leq \big( (\tilde{\bm{h}_{iik}^{\Re}})^T \bm{w_i}^{\Re} -(\tilde{\bm{h}_{iik}^{\Im} }  )^T \bm{w_i}^{\Im}
		- \sqrt{( \sigma_n^2+\sum_{j\ne i}^{N_{BS}}|| \tilde{ \bm{h}_{jik}^T}\bm{w}_j||^2)\overline {\Gamma_{ik}} } \big)  \mathrm{tan}\theta | \bm{e}_{ik} \} \geq \eta_{ik},&\\
		\mathrm{Pr}\{ -   (\tilde{\bm{h}_{iik}^{\Re}})^T \bm{w_i}^{\Im} - (\tilde{\bm{h}_{iik}^{\Im}}   )^T \bm{w_i}^{\Re}\leq \big( (\tilde{\bm{h}_{iik}^{\Re}})^T \bm{w_i}^{\Re}  -(\tilde{\bm{h}_{iik}^{\Im} }  )^T \bm{w_i}^{\Im}
		- \sqrt{( \sigma_n^2+\sum_{j\ne i}^{N_{BS}}||\tilde{  \bm{h}_{jik}^T} \bm{w}_j||^2)\overline {\Gamma_{ik}} }\big)     \mathrm{tan}\theta | \bm{e}_{ik} \} \geq \eta_{ik}.
		\end{array}  
		\right.  
		\label{eq:Partial CI prob 2}
		\end{equation}
	\end{small}%

	To decouple the inter-cell interference at each BS, we introduce two auxiliary variables $\varphi_{ik}$ and $u_{ik}$. Then,  \eqref{eq:Partial CI prob 2} is transformed as

	\begin{small}
		\begin{equation}  
		\left\{  
		\begin{array}{lr}
		(C12a):  \mathrm{Pr}\{   [ \hat{\bm{h}_{iik}^{\Im}}+\bm{e}_{ik}^{\Im}  - \hat{\bm{h}_{iik}^{\Re}}\mathrm{tan}\theta- \bm{e}_{ik}^{\Re}\mathrm{tan}\theta;    
		\hat{\bm{h}_{iik}^{\Re}}+\bm{e}_{ik}^{\Re} +\hat{\bm{h}_{iik}^{\Im}}\mathrm{tan}\theta+ \bm{e}_{ik}^{\Im}\mathrm{tan}\theta]^T  [\bm{w}_i^{\Re}; \bm{w}_i^{\Im}]  \\
		~~~~~~~~~~~~~ \leq -\sqrt{ \overline{\Gamma_{ik}}} ( \sigma_n+    \varphi_{ik} )\mathrm{tan}\theta\} \geq \eta_{ik}, &\\
		(C12b):\mathrm{Pr}\{ [ -\hat{\bm{h}_{iik}^{\Im}}-\bm{e}_{ik}^{\Im}  - \hat{\bm{h}_{iik}^{\Re}}\mathrm{tan}\theta- \bm{e}_{ik}^{\Re}\mathrm{tan}\theta;   -\hat{\bm{h}_{iik}^{\Re}}-\bm{e}_{ik}^{\Re} + \hat{\bm{h}_{iik}^{\Im}}\mathrm{tan}\theta+ \bm{e}_{ik}^{\Im}\mathrm{tan}\theta]^T  [\bm{w}_i^{\Re}; \bm{w}_i^{\Im}] \\
		~~~~~~~~~~~~~\leq -\sqrt{ \overline{\Gamma_{ik}}} ( \sigma_n+    \varphi_{ik} )\mathrm{tan}\theta \} \geq \eta_{ik},&\\
		(C12c):\varphi_{ik}^2 \geq u_{ik}, ~(C12d):  u_{ik} \geq \sum_{j \ne i}^{N_{BS}}   ||  \tilde{\bm{h}_{jik}^T}\bm{w}_j ||^2.
		\label{eq:Partial CI det P8}
		\end{array}  
		\right.  
		\end{equation} 
	\end{small}%
	
	Now we handle the constraints in \eqref{eq:Partial CI det P8} one by one.
	For constraint (C12a),  it can be equivalently written in the form as
	\begin{small}
		\begin{equation}  
		\begin{split}
		& (C12a): \overline{\bm{f}_{ik}}^T [\bm{w_i}^{\Re};\bm{w_j}^{\Im}]+\Phi^{-1}(\eta_{ik} )||   \bm{\Lambda}^{\frac{1}{2}}_{1}  [\bm{w}_i^{\Re};\bm{w}_i^{\Im}]||_2 \leq  -(\varphi_{ik}+\sigma_n)\sqrt{ \overline{  \Gamma_{ik}}}  \mathrm{tan}\theta, 
		\label{eq:C12a 1}
		\end{split}
		\end{equation}  
	\end{small}%
	where $ \overline{\bm{f}_{ik}}=  [ \hat{\bm{h}_{iik}^{\Im}} - \hat{\bm{h}_{iik}^{\Re}}\mathrm{tan}\theta;   \hat{\bm{h}_{iik}^{\Re}}+ \hat{\bm{h}_{iik}^{\Im}}\mathrm{tan}\theta]$ and $\bm{\Lambda}_{ik}=\mathrm{diag} ( \underbrace{ (1+\mathrm{tan}\theta)^2\sigma_{ik}^2,...,(1+\mathrm{tan}\theta)^2\sigma_{ik}^2}_{2M})$. According Schur complement,  constraint (C12a) can be finally given as

	
	\begin{small}
		\begin{equation}  
		\begin{split}
		&(C12a):
		\left[
		\begin{matrix}
		\frac{ \big(- \overline{\bm{f}_{ik}}^T [\bm{w}_i^{\Re};\bm{w}_i^{\Im}]-(\sigma_n+\varphi_{ik}) \sqrt{ \overline{  \Gamma_{ik}}}\mathrm{tan}\theta\big) \bm{I}}{\Phi^{-1}(\eta_{ik}) }&   \bm{\Lambda}^{\frac{1}{2}}_{ik} [\bm{w}_i^{\Re};\bm{w}_i^{\Im}]\\
		(\bm{\Lambda}^{\frac{1}{2}}_{ik} [\bm{w}_i^{\Re};\bm{w}_i^{\Im}]) ^T& \frac{-\overline{\bm{f}_{ik}}^T [\bm{w}_i^{\Re};\bm{w}_i^{\Im}]-(\sigma_n+\varphi_{ik})\sqrt{ \overline{  \Gamma_{ik}}}\mathrm{tan}\theta}{\Phi^{-1}(\eta_{ik}) } \\
		\end{matrix}
		\right]  
		\succeq \bm{0}.
		\label{eq:Partial CI C12c}
		\end{split}
		\end{equation}  
	\end{small}%
	Similarly,  constraint (C12b) can be directly written as


	
	\begin{small}
		\begin{equation}  
		\begin{split}
		&(C12b):\left[
		\begin{smallmatrix}
		\frac{ \big(- \overline{\bm{d}_{ik}}^T [\bm{w}_i^{\Re};\bm{w}_i^{\Im}]-(\sigma_n+\varphi_{ik}) \sqrt{ \overline{  \Gamma_{ik}}}\mathrm{tan}\theta\big) \bm{I}}{\Phi^{-1}(\eta_{ik}) }&   \bm{\Lambda}^{\frac{1}{2}}_{ik} [\bm{w}_i^{\Re};\bm{w}_i^{\Im}]\\
		(\bm{\Lambda}^{\frac{1}{2}}_{ik} [\bm{w}_i^{\Re};\bm{w}_i^{\Im}]) ^T& \frac{-\overline{\bm{d}_{ik}}^T [\bm{w}_i^{\Re};\bm{w}_i^{\Im}]-(\sigma_n+\varphi_{ik})\sqrt{ \overline{  \Gamma_{ik}}}\mathrm{tan}\theta}{\Phi^{-1}(\eta_{ik}) } 
		\end{smallmatrix}
		\right] 
		\succeq \bm{0},
		\label{eq:C12d}
		\end{split}
		\end{equation}  
	\end{small}%
	where $\overline{\bm{d}_{ik}}=[- \hat{ \bm{h}_{iik}^{\Im}}-\hat{ \bm{h}_{iik}^{\Re}} \mathrm{tan}\theta;-\hat{ \bm{h}_{iik}^{\Re}}+\hat{ \bm{h}_{iik}^{\Im}} \mathrm{tan}\theta]$. Now we turn to handle the constraint (C12c). By stacking the variables $\varphi_{ik}$ and $u_{ik}$ into a long vector that $\bm{t}_{ik}=[\varphi_{ik},u_{ik}]^T$ and defining $\bm{S}_{ik}=\bm{t}_{ik}\bm{t}_{ik}^H$, (C12c) can be relaxed in the form as 
	
	\begin{small}
		\begin{equation}
		\begin{split}
		&(C12c):\mathrm{Tr}(\bm{\Pi}\bm{S}_{ik})+\bm{l}\bm{t}_{ik} \geq 0; \mathrm{and}
		\left[
		\begin{matrix}
		\bm{S}_{ik},& \bm{t}_{ik}\\
		(\bm{t}_{ik})^T,& 1
		\end{matrix}
		\right]
		\succeq \bm{0},
		\label{eq:Partial CI C12a}
		\end{split}
		\end{equation}
	\end{small}%
	where $\bm{\Pi}=	\left[
	\begin{matrix}
	1,& 0\\
	0,& 0
	\end{matrix}
	\right]$ and vector $ \bm{l}=[0,-1]$.
	Next we can handle constraint (C12d), which can be further written as

	\begin{small}
		\begin{equation}  
		\begin{split}
		&(C12d): \sum_{j \ne i}^{N_{BS}}\big( \bm{e}_{ik}^T\bm{W}_j \bm{e}_{ik}^{\ast})+\sum_{j \ne i}^{N_{BS}} \big(\bm{e}_{ik}^T\bm{W}_j \hat{\bm{h}_{jik}^{\ast}}\big) +\sum_{j \ne i}^{N_{BS}}\big(\hat{\bm{h}_{jik}^T}  \bm{W}_j \bm{e}_{ik}^{\ast}\big)+\sum_{j \ne i}^{N_{BS}} \big(\bm{h}_{jik}^T \bm{W}_j \bm{h}_{jik}^{\ast} \big)-u_{ik} \leq 0.
		\label{eq:Partial CI det C14b}
		\end{split}
		\end{equation}  
	\end{small}%
	
	To handle the inequality with S-procedure, we need to construct another premise $\bm{e}_{ik}^T \bm{I}_{M} \bm{e}_{ik}^{\ast} \leq \nu^2$ to guarantee \eqref{eq:Partial CI det C14b} always hold. We now calculate the  value of  $\nu^2$ according to Lemma 3.

	\textbf{\textit{Lemma 3}} (links $\nu^2$ to CSI error variance $\sigma_{ik}^2$): 
	Provided the element of CSI error follows normal distribution such that $[\bm{e}_{ik}]_m \sim \mathcal{CN}(0,\sigma_{ik}^2)$, the term $\bm{e}_{ik}^T \bm{I}_{M} \bm{e}_{ik}^{\ast}$ is bounded by $\nu^2=\Upsilon^{-1}(\delta) \sigma_{ik}^2$. Explicitly, $\delta$ physically represents the outage probability of $\bm{e}_{ik}^T \bm{I}_{M} \bm{e}_{ik}^{\ast}$ violating its upper bound,   and can be set to approach  1, i.e.  $\delta=0.99$ \cite{Nasseri2016Chance}.  $\Upsilon^{-1}(\cdot)$ is the inverse function of the cdf of a chi-square distributed variable with $M$ degree of freedom.

	Proof: Please see Appendix B. $\blacksquare$
	
	Now with the known value of $\nu^2$, the implication $\big( \bm{e}_{ik}^T \bm{I}_{M} \bm{e}_{ik}^{\ast} \big)- \nu^2 \leq 0 \Rightarrow$ (\ref{eq:Partial CI det C14b}) holds if and only if the following LMI holds


	\begin{small}
		\begin{equation}
		\begin{split}
		&(C12d):\left[
		\begin{matrix}
		\phi_{ik} \bm{I}_{M}- \sum_{j\ne i}^{N_{BS}} \bm{W}_j ,&-\sum_{j \ne i}^{N_{BS}}  \hat{\bm{h}_{jik}^{T}} \bm{W}_j  \\
		-(\sum_{j \ne i}^{N_{BS}} \hat{\bm{h}_{jik}^{T}} \bm{W}_j )^H,& -\phi_{ik}\nu^2-\sum_{j \ne i}^{N_{BS}}(\hat{\bm{h}_{jik}^T} \bm{W}_j \hat{\bm{h}_{jik}^{\ast}} )+u_{ik}
		\end{matrix}
		\right]  \succeq \bm{0}, ~\mathrm{and}~ \phi_{ik} \geq 0.
		\label{eq:Partial CI prob C12d}
		\end{split}
		\end{equation}
	\end{small}%

	Now, the original constraint (C12) is transformed into equivalent constraints (C12a)-(C12d), as shown by Eqs. \eqref{eq:Partial CI C12c}, \eqref{eq:C12d}, \eqref{eq:Partial CI C12a} and \eqref{eq:Partial CI prob C12d}.
	The optimization problem becomes

	\begin{small}
		\begin{equation}
		\begin{split}
		& P6~ (\mathrm{Partial-CI -Prob}):\operatorname*{argmin}\limits_{\bm{w}_j, \forall j \in N_{BS}}   \sum_{j=1}^{N_{BS}} \mathrm{Tr}( \bm{W}_j ),\\
		&~\mathrm{s.t}~(C11):  \mathrm{Tr}(\bm{W}_j  ) \leq P_{max}, \forall j\in N_{BS},~(C12a): \eqref{eq:Partial CI C12c},~\\
		&~~~~~(C12b): \eqref{eq:C12d},~(C12c): \eqref{eq:Partial CI C12a},~(C12d): \eqref{eq:Partial CI prob C12d},~(C13):  \left[
		\begin{matrix}
		\bm{W}_j& \bm{w}_j\\
		\bm{w}_j^T& 1
		\end{matrix}
		\right] \succeq 0,\forall j \in N_{BS},
		\label{eq:unknown_P1}
		\end{split}
		\end{equation}
	\end{small}%
	which is readily solved by CVX.

	\subsection{Deterministic Optimization}
	
	We now solve the problem in terms of deterministic manner. Accordingly, the problem formulation is given as

	\begin{small}
		\begin{equation}
		\begin{split}
		& P7~ (\mathrm{Partial-CI -Det}): \operatorname*{argmin}\limits_{\bm{w}_j, \forall j\in N_{BS}}  \sum_{j=1}^{N_{BS}} \left \| w_j  \right \|    ^2 ,\\
		&~\mathrm{s.t~}(C14): \left \| w_j  \right \|    ^2 \leq p_{max}, \forall j\in N_{BS},~(C15):  \operatorname*{min}\limits_{\bm{e_{ik}\in \Delta}} \Gamma_{ik} \geq \overline{ \Gamma_{ik} },  \forall i \in N_{BS}, k \in K.
		\label{eq:Partial CI det P6}
		\end{split}
		\end{equation}
	\end{small}%
	
	To make the optimization problem solvable, we need to process constraint (C15) that contains infinite possibilities. Similarly, we introduce two $u_{ik}$ and $\varphi_{ik}$ to decouple the inter-cell interference. Hence, (C15) equals to

	\begin{small}
		\begin{equation}  
		\begin{split}
		\left\{  
		\begin{array}{lr}
		(C15a):  \operatorname*{min}\limits_{\bm{e_{ik}\in \Delta}} [ \hat{\bm{h}_{iik}^{\Im}}+\bm{e}_{ik}^{\Im}  - \hat{\bm{h}_{iik}^{\Re}}\mathrm{tan}\theta- \bm{e}_{ik}^{\Re}\mathrm{tan}\theta;  \hat{\bm{h}_{iik}^{\Re}}+\bm{e}_{ik}^{\Re}   \\
		~~~~~~~~~~~~~+ \hat{\bm{h}_{iik}^{\Im}}\mathrm{tan}\theta+ \bm{e}_{ik}^{\Im}\mathrm{tan}\theta]^T [\bm{w}_i^{\Re}; \bm{w}_i^{\Im}]  \leq -\sqrt{ \overline{\Gamma_{ik}}} ( \sigma_n+    \varphi_{ik} )\mathrm{tan}\theta    , &\\
		(C15b): \operatorname*{min}\limits_{\bm{e_{ik}\in \Delta}} [- \hat{\bm{h}_{iik}^{\Im}}-\bm{e}_{ik}^{\Im}  - \hat{\bm{h}_{iik}^{\Re}}\mathrm{tan}\theta- \bm{e}_{ik}^{\Re}\mathrm{tan}\theta;  -\hat{\bm{h}_{iik}^{\Re}}- \bm{e}_{ik}^{\Re}   \\ 
		~~~~~~~~~~~~~+\hat{\bm{h}_{iik}^{\Im}}\mathrm{tan}\theta+ \bm{e}_{ik}^{\Im}\mathrm{tan}\theta]^T [\bm{w}_i^{\Re}; \bm{w}_i^{\Im}]  \leq -\sqrt{ \overline{\Gamma_{ik}}} ( \sigma_n+    \varphi_{ik} )\mathrm{tan}\theta,&\\
		(C15c):\varphi_{ik}^2 \geq u_{ik},~(C15d): u_{ik} \geq \sum_{j \ne i}^{N_{BS}}   ||  \bm{h}_{jik}\bm{w}_j ||^2,
		\label{eq:Partial CI det C14}
		\end{array}  
		\right.  
		\end{split}
		\end{equation} 
	\end{small}%
	
	Firstly, constraint (C15a) can be written as 
	\begin{small}
		\begin{equation}  
		\begin{split}
		(C15a):[  \bm{e}_{ik}^{\Im}-\bm{e}_{ik}^{\Re}\mathrm{tan\theta}; \bm{e}_{ik}^{\Re}+\bm{e}_{ik}^{\Im}\mathrm{tan\theta}    ]^T [ \bm{w}_i^{\Re}, \bm{w}_i^{\Im}]+\rho_{ik} \leq 0,
		\label{eq:Partial CI C8 1}
		\end{split}
		\end{equation}  
	\end{small}%
	where $\varrho_{ik}=  (\hat{\bm{h}_{jik}^{\Im}})^T  \bm{w}_j^{\Re}-(\hat{\bm{h}_{jik}^{\Re}})^T  \bm{w}_j^{\Re} \mathrm{tan\theta}+(\hat{\bm{h}_{jik}^{\Re}} )^T \bm{w}_j^{\Im}+(\hat{\bm{h}_{jik}^{\Im}})^T  \bm{w}_j^{\Im} \mathrm{tan\theta}+(\varphi_{ik}+\sigma_n) \sqrt{\overline{\Gamma_{ik}}}\mathrm{tan\theta}$.  To hold the implication   $[\sqrt{\bm{e}_{ik}^{\Im} -\bm{e}_{ik}^{\Re}\mathrm{tan}\theta};\sqrt{\bm{e}_{ik}^{\Re} +\bm{e}_{ik}^{\Im}\mathrm{tan}\theta  }] ^T\bm{I}_{2N}$ $[\sqrt{\bm{e}_{ik}^{\Im} -\bm{e}_{ik}^{\Re}\mathrm{tan}\theta};\sqrt{\bm{e}_{ik}^{\Re} +\bm{e}_{ik}^{\Im}\mathrm{tan}\theta  }]- \rho^2 \leq 0 \Rightarrow$ $
	[\sqrt{\bm{e}_{ik}^{\Im} -\bm{e}_{ik}^{\Re}\mathrm{tan}\theta};\sqrt{\bm{e}_{ik}^{\Re} +\bm{e}_{ik}^{\Im}\mathrm{tan}\theta  }] ^T \mathrm{diag}(\bm{w}_i^{\Re};\bm{w}_i^{\Im})
	[\sqrt{\bm{e}_{ik}^{\Im} -\bm{e}_{ik}^{\Re}\mathrm{tan}\theta};\sqrt{\bm{e}_{ik}^{\Re} +\bm{e}_{ik}^{\Im}\mathrm{tan}\theta  }] 
	+ \varrho_{ik} \leq 0$, we have to guarantee the following LMI hold
	
	\begin{small}
		\begin{equation}
		\begin{split}
		&(C15a):
		\left[
		\begin{matrix}
		\varsigma_{ik} \bm{I}_{2N}-\mathrm{diag}(\bm{w}_i^{\Re};\bm{w}_i^{\Im}),& \bm{0}\\
		\bm{0},&- \varsigma_{ik} \rho^2-\varrho_{ik}
		\end{matrix}
		\right]
		\succeq \bm{0},~\mathrm{and}~ \varsigma_{ik} \geq 0,
		\label{eq:Partial CI det C14c}
		\end{split}
		\end{equation}
	\end{small}%
	where $\rho^2=\Phi^{-1}(\delta)\sqrt{  M(1+\mathrm{tan}^2\theta)      }\sigma_{ik}$  similarly calculated by Lemma 2. Also, constraint (C15b) can be written as
	\begin{small}
		\begin{equation}
		\begin{split}
		&(C15b):
		\left[
		\begin{matrix}
		\theta_{ik} \bm{I}_{2N}-\mathrm{diag}(\bm{w}_i^{\Re};\bm{w}_i^{\Im}),& \bm{0}\\
		\bm{0},&- \theta_{ik} \rho^2-\beta_{ik}
		\end{matrix}
		\right]
		\succeq \bm{0},~\mathrm{and}~ \theta_{ik} \geq 0,
		\label{eq:Partial CI det C14d}
		\end{split}
		\end{equation}
	\end{small}%
	where $\beta_{ik}= [-\hat{h_{ijk}^{\Im}} \bm{w}_j^{\Re}-\hat{h_{ijk}^{\Re}}  \mathrm{tan\theta}\bm{w}_j^{\Re}-\hat{h_{ijk}^{\Re}} \bm{w}_j^{\Im}+\hat{h_{ijk}^{\Im}} \bm{w}_j^{\Im} \mathrm{tan\theta}+(\sigma_n+\varphi_{ik})\sqrt{\overline{\Gamma_{ik}}}\mathrm{tan}\theta$.
	On the other hand, since (C15c) and (C15d) have similar structures with (C12c) and (C12d), (C15c) and (C15d) can be written in the forms as

	\begin{small}
		\begin{equation}
		\begin{split}
		&(C15c):\mathrm{Tr}(\bm{\Pi}\bm{S}_{ik})+\bm{l}\bm{t}_{ik} \geq 0; \mathrm{and}
		\left[
		\begin{matrix}
		\bm{S}_{ik},& \bm{t}_{ik}\\
		(\bm{t}_{ik})^T,& 1
		\end{matrix}
		\right]
		\succeq \bm{0},
		\label{eq:Partial CI det C14a}
		\end{split}
		\end{equation}
	\end{small}%

	\begin{small}
		\begin{equation}
		\begin{split}
		&(C15d):
		\left[
		\begin{matrix}
		-\phi_{ik} \bm{I}_{M}- \sum_{j\ne i}^{N_{BS}} \bm{W}_j ,&-\sum_{j \ne i}^{N_{BS}}  \bm{W}_j\hat{\bm{h}_{jik}^{\ast}}  \\
		-(\sum_{j \ne i}^{N_{BS}} \bm{W}_j\hat{\bm{h}_{jik}^{\ast}} )^H,& -\phi_{ik}\nu^2-\sum_{j \ne i}^{N_{BS}}(\hat{\bm{h}_{jik}} \bm{W}_j \hat{\bm{h}_{jik}^H} )+u_{ik}
		\end{matrix}
		\right] 
		\succeq \bm{0}, ~\mathrm{and}~ \phi_{ik} \geq 0.
		\label{eq:Partial CI det C14b1}
		\end{split}
		\end{equation}
	\end{small}%

	Now, the problem is transformed into 
	\begin{small}
		\begin{equation}
		\begin{split}
		& P8~ (\mathrm{Partial-CI -Det}): \operatorname*{argmin}\limits_{\bm{w}_j, \forall j\in N_{BS}}  \sum_{j=1}^{N_{BS}} \left \| w_j  \right \|    ^2 ,\\
		&~~\mathrm{s.t~}(C14): \mathrm{Tr}(\bm{W}_j)  \leq p_{max}, \forall j\in N_{BS},~(C15a):\eqref{eq:Partial CI det C14c}, ~(C15b):\eqref{eq:Partial CI det C14d}, \forall i \in N_{BS}, k \in K,\\
		& ~~~~~~~(C15c):  \eqref{eq:Partial CI det C14a}, ~(C15d):\eqref{eq:Partial CI det C14b1},~(C16):  \left[
		\begin{matrix}
		\bm{W}_j& \bm{w}_j\\
		\bm{w}_j^T& 1
		\end{matrix}
		\right] \succeq 0,\forall j \in N_{BS},
		\label{eq:Partial CI det P7}
		\end{split}
		\end{equation}
	\end{small}%
	which is readily solved by CVX.

	\section{ Statistically-Coordinated CI Precoding   }
	
	In Section IV,  instantaneous CSI needs to be shared among the BSs to facilitate the optimization. In this section, to further reduce the coordination overhead, we propose a  CI precoding scheme that BSs need to know its local estimated CSI while only sharing statistical inter-cell channel with others. Hence, the coordination overhead can be significantly reduced compared to the above instantaneous CSI based schemes.
	Let us define $\bm{R}_{jik}$ as the correlation matrix of BS $j$ to the user $\mathrm{U_{ik}}$. 
	Assuming white channel covariance \cite{Liao2011QoS}, the true channel correlation matrix can be written as $\bm{R}_{jik}=\bm{u}_{jik}^H\bm{u}_{jik}+ \sigma ^2 \bm{I}_{M},\forall j \ne i, j\in N_{BS}$, where  $\bm{u}_{jik}$ denotes the mean and $\bm{\Sigma}_{jik}$ denotes the covariance matrix of channel $\bm{h}_{jik}$. Accordingly, the problem is formulated as


	\begin{small}
		\begin{equation}
		\begin{split}
		& P9~ (\mathrm{Stat-CI }): \operatorname*{argmin}\limits_{\bm{w}_j, \forall j\in N_{BS}}  \sum_{j=1}^{N_{BS}} \left \| w_j  \right \|    ^2 ,\\
		&~~\mathrm{s.t~}(C20): \left \| w_j  \right \|    ^2 \leq p_{max}, \forall j\in N_{BS},~(C21):  \operatorname*{min}\limits_{\bm{e_{ik}\in \Delta}} \Gamma_{ik} \geq \overline{ \Gamma_{ik} },  \forall i \in N_{BS}, k \in K.
		\label{eq:Stat CI P10}
		\end{split}
		\end{equation}
	\end{small}%

	We now need to handle  constraint (C21), which can be equivalently written as

	\begin{small}
		\begin{equation}  
		\begin{split}
		\left\{  
		\begin{array}{lr}
		(C21c):  \operatorname*{min}\limits_{\bm{e_{ik}\in \Delta}} [ \hat{\bm{h}_{iik}^{\Im}}+\bm{e}_{ik}^{\Im}  - \hat{\bm{h}_{iik}^{\Re}}\mathrm{tan}\theta- \bm{e}_{ik}^{\Re}\mathrm{tan}\theta;  \hat{\bm{h}_{iik}^{\Re}}+\bm{e}_{ik}^{\Re}   \\ 
		~~~~~~~~~~~~~~~~~~~~+ \hat{\bm{h}_{iik}^{\Im}}\mathrm{tan}\theta+ \bm{e}_{ik}^{\Im}\mathrm{tan}\theta]^T [\bm{w}_i^{\Re}; \bm{w}_i^{\Im}]  \leq -\sqrt{ \overline{\Gamma_{ik}}} ( \sigma_n+    \varphi_{ik} )\mathrm{tan}\theta    , &\\
		(C21d): \operatorname*{min}\limits_{\bm{e_{ik}\in \Delta}} [- \hat{\bm{h}_{iik}^{\Im}}-\bm{e}_{ik}^{\Im}  - \hat{\bm{h}_{iik}^{\Re}}\mathrm{tan}\theta- \bm{e}_{ik}^{\Re}\mathrm{tan}\theta;  -\hat{\bm{h}_{iik}^{\Re}}-\bm{e}_{ik}^{\Re}   \\ 
		~~~~~~~~~~~~~~~~~~~~+\hat{\bm{h}_{iik}^{\Im}}\mathrm{tan}\theta+ \bm{e}_{ik}^{\Im}\mathrm{tan}\theta]^T [\bm{w}_i^{\Re}; \bm{w}_i^{\Im}]  \leq -\sqrt{ \overline{\Gamma_{ik}}} ( \sigma_n+    \varphi_{ik} )\mathrm{tan}\theta,&\\
		(C21c):\varphi_{ik}^2 \geq u_{ik}, 
		(C21d):\sum_{j \ne i}^{N_{BS}} \mathrm{Tr}(   \bm{w}_j\bm{w}_j^H \bm{R}_{jik}  ) \leq   u_{ik}.
		\label{eq:Partial CI det C21}
		\end{array}  
		\right.  
		\end{split}
		\end{equation} 
	\end{small}%
	As can be seen, constraint (C21d) itself is a LMI while (C21a), (C21b) and (C21c) can similarly be solved as presented in Section IV. Hence, the problem can be re-formulated as

	\begin{small}
		\begin{equation}
		\begin{split}
		& P10~ (\mathrm{Stat-CI}):\operatorname*{argmin}\limits_{\bm{W}_j, \forall j \in N_{BS}}   \sum_{j=1}^{N_{BS}} \mathrm{Tr}( \bm{W}_j ),\\
		&~\mathrm{s.t}~(C20):  \mathrm{Tr}(\bm{W}_j  ) \leq P_{max}, \forall j\in N_{BS},~(C21a): \eqref{eq:Partial CI det C14c},~(C22b): \eqref{eq:Partial CI det C14d},~(C22c): \eqref{eq:Partial CI det C14a},\\
		&~~~~~~(C21d):\sum_{j \ne i}^{N_{BS}} \mathrm{Tr}(   \bm{w}_j\bm{w}_j^H \bm{R}_{jik}  ) \leq   u_{ik} ,~(C22):  \left[
		\begin{matrix}
		\bm{W}_j& \bm{w}_j\\
		\bm{w}_j^T& 1
		\end{matrix}
		\right] \succeq 0,\forall j \in N_{BS},
		\label{eq:Stat CI P9}
		\end{split}
		\end{equation}
	\end{small}%
	which can be readily solved by CVX.

	\section{Complexity and Overhead Analysis}
	
	In this section, we present coordination overhead and complexities of different algorithms, respectively. The associated analytical expressions are summarized in Table I.

	\subsection{Coordination Overhead}
	By the Full-CI-Prob/Det algorithms, each BS needs to share CSI and the intended transmitted symbols to other coordinated BSs. With $K$ users in each BS and $N$ coordinated BSs, the total overhead of sharing CSI and data is given $\mathcal{O}\big(N(N-1)K  (N\chi   _{C}+\chi_{S}) \big)$, where $ \chi   _{C}$ denotes the required bits for describing one user's CSI while $ \chi   _{S}$ denotes the required bits for exchanging symbols. Evidently, the  Full-CI-Prob/Det algorithms show the same overhead with conventional CoMP systems, which are suited for the scenario where the coordinated BSs are connected with high-capacity and delay-free links. 
	By the Partial-CI-Prob/Det algorithms, only CSI needs to be shared among the BSs. Hence, the overhead is significantly reduced to $\mathcal{O}(N^2(N-1)K  \chi   _{C})$. 
	By the Stat-CI algorithm, BSs only share statistical CSI of inter-cell channel   with others, and the required bits can be ignored compared to other  instantaneous CSI acquisition-based schemes.

	\subsection{Computational Complexity}
	Now we analyze the complexities in each algorithms. For the interior-point methods based solver, the overall complexity can be given as $\mathrm{ln}(\frac{1}{\epsilon})\sqrt{c_{b}}(c_{f}+c_{g})$ \cite{Wang2014Outage}. Specifically, $\mathrm{ln}(\frac{1}{\epsilon})$ relates to the accuracy setup. $\sqrt{c_{b}}$ represents the barrier parameter measuring
	the geometric complexity of the conic constraints. $c_{f}$ and $c_{g}$ represent the complexities cost on forming and factorization of $n\times n$ matrix of the optimization problem \footnote{By the interior-point methods based solver, a search direction is found by solving a system of linear equations in $n$ unknowns. $c_{f}$ is calculated as $c_{f}=n\sum_{j=1}^{P}k_j^3+n^2\sum_{j=1}^{P}k_j^2+n\sum_{j=P+1}^{m}k_j^2$, where $k_j$ presents the size of the $j$-th constraint. Specifically, the terms $n\sum_{j=1}^{P}k_j^3+n^2\sum_{j=1}^{P}k_j^2$ come from $P$ LMI constraints while the term $n\sum_{j=P+1}^{m}k_j^2$ comes from $m-P$ second order cone constraints in the problem formulation. $c_{g}$ is calculated as $c_{g}=n^3$ (eq. (18) \cite{Wang2014Outage}). }.  
	For the first scheme, P2 handles the optimization in a probabilistic manner. 
	It involves $N_{BS}$ LMI (trace) in $(C1)$ of size 1. 
	$2N_{BS}K$ LMI inequalities in $(C2)$ of size $2N_{BS}+1$.
	$N_{BS}K$ LMI inequalities in $(C3)$ of size $M+1$. Hence, barrier parameter is given as $\beta_{1}=\sqrt{N_{BS}(MK+3K+1+4KN_{BS})} $.
	On the other hand, P4 handles the problem in a deterministic manner. 
	It involves $N_{BS}$ LMI (trace) in $(C7)$ of size 1. 
	$2N_{BS}K$ LMI inequalities in $(C8)$ of size $2N_{BS}+1$. 
	$N_{BS}K$ linear inequality in (C9). 
	$N_{BS}K$ linear inequality in (C10). 
	$N_{BS}K$ LMI inequalities in $(C10)$ of size $M+1$. Hence, its barrier parameter is given as  $ \beta_{2}=\sqrt{N_{BS}(5+4N_{BS}+KM+K)} $.
	
	In the second scheme, P6 first handles the problem in a probabilistic manner. 
	It involves $N_{BS}$ LMI (trace) in $(C11)$ of size 1. 
	$N_{BS}K$ LMI inequalities  of size $2M+1$ in $(C12a)$.
	$N_{BS}K$ LMI inequalities  of size $2M+1$ in $(C12b)$.  
	$N_{BS}K$ LMI (trace)  of size 1 and $N_{BS}K$ LMI inequalities of size 3 in $(C12c)$.
	$N_{BS}K$ LMI inequalities of size $(N_{BS}-1)M^2+1$ and $N_{BS}K$ linear constraints in $(C12d)$.
	$N_{BS}K$ LMI inequalities of size $M+1$ in $(C13)$. Hence, its barrier parameter is given as  $\beta_{3}=\sqrt{N_{BS}(1+5K+2K(3M+2)+K((N_{BS}-1)M^2+1))}$.
	On the other hand, P8 handles the problem in a deterministic manner. It involves 
	$N_{BS}$ LMI (trace) in $(C14)$ of size 1. 
	$N_{BS}K$ LMI inequalities of size $2M+1$  and $N_{BS}K$ linear constraints in $(C15a)$. 
	$N_{BS}K$ LMI inequalities  of size $2M+1$ and $N_{BS}K$ linear constraints in $(C15b)$. 
	$N_{BS}K$ LMI (trace) in $(C15c)$ of size 1 and $N_{BS}K$ LMI inequalities in $(C15c)$ of size 3. 
	$N_{BS}K$ LMI inequalities of size $M+1$ and $N_{BS}K$ linear constraints in $(C15d)$.
	$N_{BS}K$ LMI inequalities in $(C16)$ of size $M+1$. Hence, its barrier parameter is given as  $ \beta_{4}=\sqrt{  N_{BS}(2+M+K(8+5M+2))         }$.
	
	Now we check the complexity of the Stat-CI algorithm. (P10) involves 
	$N_{BS}$ LMI (trace)  of size 1 in (C20).
	$N_{BS}K$ inequalities of size $2M+1$ and $N_{BS}K$ linear constraints in (C21a).
	$N_{BS}K$ inequalities of size $2M+1$ and $N_{BS}K$ linear constraints in (C21b).
	$N_{BS}K$ LMI (trace)  of size 1 and $N_{BS}K$ LMI inequalities of size 3 in (C21c).
	$N_{BS}K$ LMI (trace)  of size 1 in (C21d).
	$N_{BS}$ LMI inequalities of size $M+1$ in (C22). Hence, its barrier parameter is given as $\beta_{5}=\sqrt{ N_{BS} (2+9K+4KM+M )  } $.


	\begin{table*}
		\begin{center}
			\label{tab:simsetup}
			\centerline{TABLE \uppercase\expandafter{\romannumeral 1.} Overhead and complexity analysis, with accuracy factor $\epsilon$}
			\begin{tabular}{|c|c|c|} 
				\hline
				Algorithms&Overhead& Complexity\\
				\hline
				Full-CI-Prob& $\mathcal{O}(N^2(N-1)K  \chi   _{C}$& $ln(\frac{1}{\epsilon})\beta_{1} \big(   n(N+2N_{BS}K(2N_{BS}+1)^3+N_{BS}K(M+1)^3) $\\	
				&         $+    N(N-1)K  \chi   _{S})  $                                                         & $+n^2( N+2N_{BS}K(2N_{BS}+1)^2+N_{BS}K(M+1)^2  )        +n^3                     \big)$\\
				\hline
				Full-CI-Det  & $\mathcal{O}(N^2(N-1)K  \chi   _{C}$& $ln(\frac{1}{\epsilon})\beta_{2}\big(  n(  N_{BS} +2N_{BS}K(2N_{BS}+1)^3+2N_{BS}K+N_{BS}K(M+1)^3 )$\\	
				&         $+    N(N-1)K  \chi   _{S})  $                                 &$+n^2(  N_{BS} +2N_{BS}K(2N_{BS}+1)^2+2N_{BS}K+N_{BS}K(M+1)^2          ) +n^3     \big)$\\
				\hline
				Partial-CI-Prob& $\mathcal{O}(N^2(N-1)K  \chi   _{C})$   & $ln(\frac{1}{\epsilon})\beta_{3} \big(n(N_{BS}+10N_{BS}K+N_{BS}K( (N_{BS}-1)M^2+1         )^3$    \\	
				
				&                                                           &$+2N_{BS}K(2M+1)^3+N_{BS}K(M+1)^3   ) +n^2(  N_{BS}+10N_{BS}K+$\\
				&                                                         & $N_{BS}K( (N_{BS}-1)M^2+1   )^2+2N_{BS}K(2M+1)^2+N_{BS}K(M+1)^2  )    +n^3      \big)$\\
				\hline
				Partial-CI-Det  & $\mathcal{O}(N^2(N-1)K  \chi   _{C})$ & $ln(\frac{1}{\epsilon}) \beta_{4}\big(  n(      N_{BS}+31N_{BS}K+2N_{BS}K(M+1)^3+2N_{BS}K(2M+1)^3    )       $\\
				&                                                           & $+n^2( N_{BS}+13N_{BS}K+2N_{BS}K(M+1)^2+2N_{BS}K(2M+1)^2   )+n^3 \big) $  \\
				\hline
				Stat-CI   &  Negligible                                                        &$ln(\frac{1}{\epsilon}) \beta_{5}\big(  n( N_{BS}+31N_{BS}K+2N_{BS}K(2M+1)^3+  N_{BS}K(M+1)^3   ) +  $\\
				&                                                             & $N^2(    N_{BS}+13N_{BS}K+2N_{BS}K(2M+1)^2+  N_{BS}K(M+1)^2   )  +n^3             \big)$\\
				\hline
			\end{tabular}
		\end{center}
	\end{table*}

	\section{Simulation Results}

\begin{figure}
		\centering
		\includegraphics[width=2.1 in]{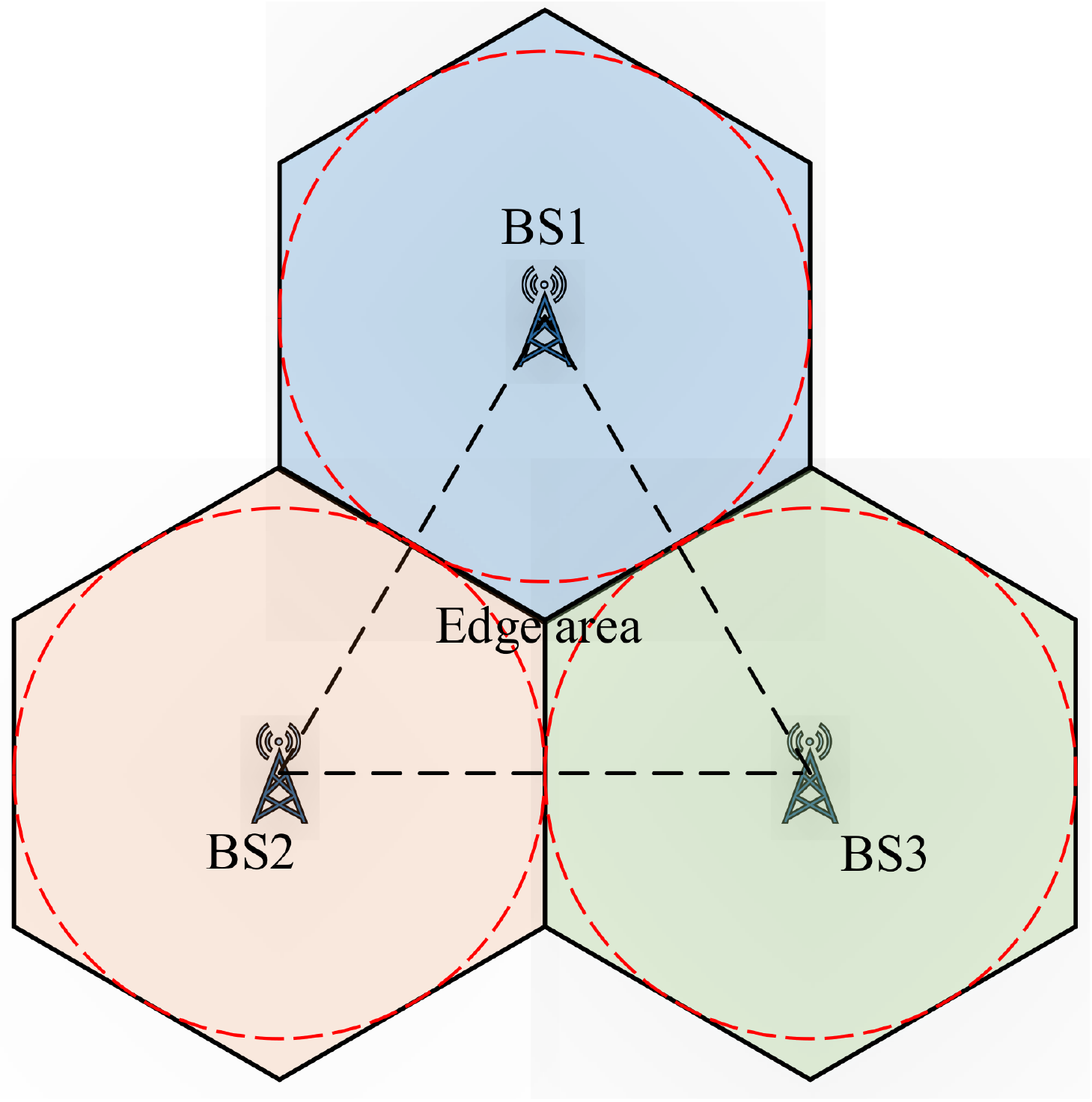}
		\caption{Illustration of system model, where 3 BSs exchange data or/and CSI for different level of multi-cell coordination.}
		\label{fig:systemmodel}
	\end{figure}

	We present the simulated performance in this section. The central frequency is set to 2 GHz with 1 MHz bandwidth. The AWGN power spectral density is -174 dBm/Hz. 
	In line with the closely relevant works \cite{Nasseri2016Chance} \cite{Shen2012Distributed}, a 3-cell coordination scenario is considered, and each cell is covered by a BS located in the center, as shown in Fig. \ref{fig:systemmodel}. The number of antennas at each BS is set to $M=4$.  The users are randomly distributed across the map with the exception of the simulation results in  Fig. \ref{fig:Cooperation20181121}, where all the users are placed in the edge area to highlight the working modes of the coordinated BSs by the different schemes. The maximum transmission power $p_{\mathrm{max}}= 100$ W, and PL model in \cite{Nasseri2016Chance} is adopted.
	The following schemes are selected as benchmarks. a) CoMP with perfect CSI \cite{Lee2012Coordinated}. Since network-level MISO is achieved by exchanging the intended transmitted symbols and CSI,  CoMP system with perfect CSI serves as a transmit power lower bound of existing schemes. b) CBF with probabilistic optimization (CBF-Prob) \cite{Nasseri2016Chance}, where the BSs  coordinate in a CBF way and probabilistic optimization is applied. c)   CBF with deterministic optimization (CBF-Det) \cite{Shen2012Distributed}, where the BSs  coordinate in a CBF way and deterministic optimization is applied.

	\begin{figure}
		\centering
		\includegraphics[width=3.0 in]{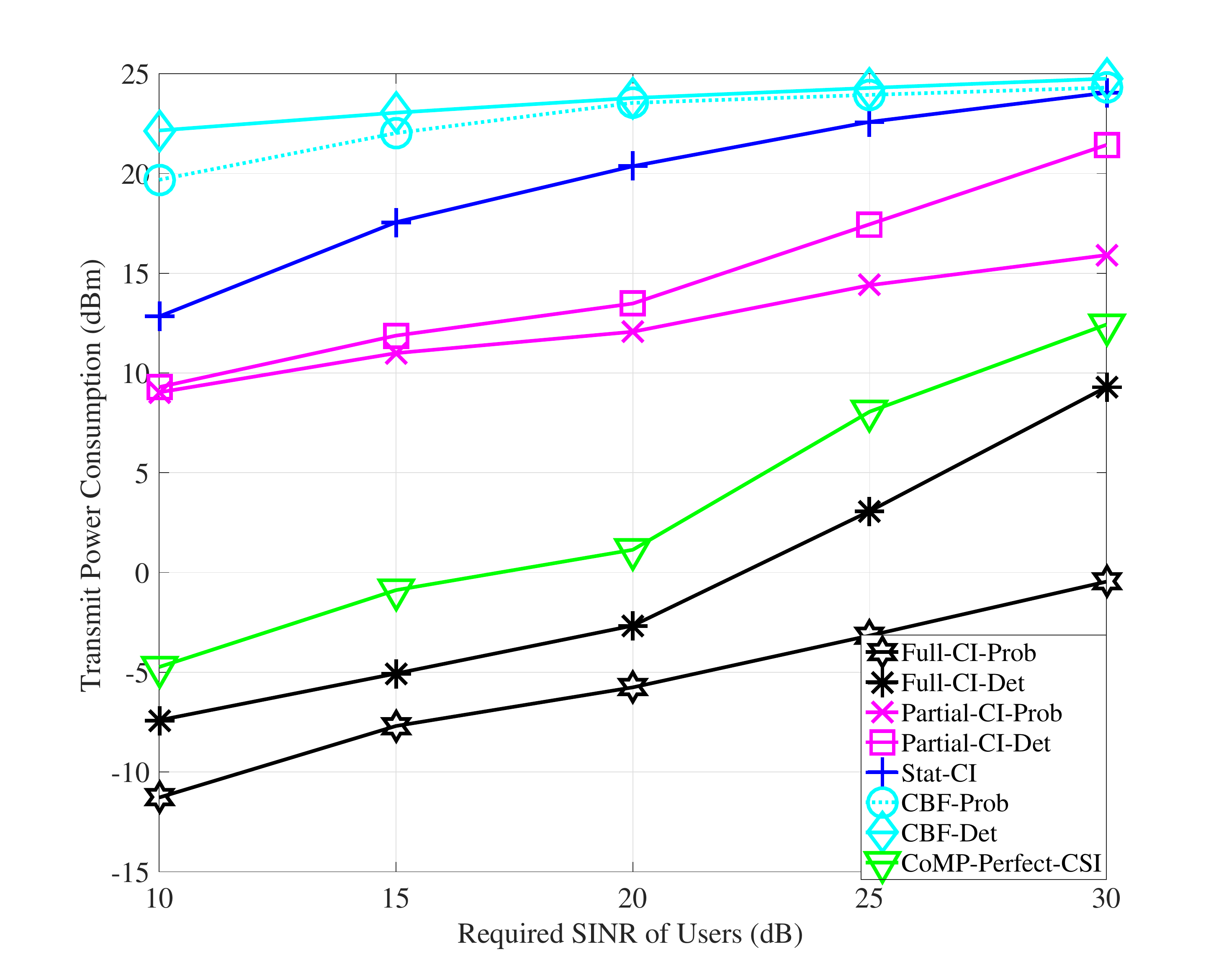}
		\caption{Impact of the users' SINR requirements $\overline{\Gamma_{ik}}$ on the total power consumption, with $K=3$ in each cell and  $\sigma_{ik}=10^{-2}$. }
		\label{fig:SINRvsPower20181204}
	\end{figure}
	
	Fig. \ref{fig:SINRvsPower20181204} shows the impact of users' SINR requirements $\overline{\Gamma_{ik}}$ on the total power consumption. It can be seen that the Full-CI-Prob and Full-CI-Det algorithms consume the lowest power. It is because the Full-CI-Prob/Det utilize both inter-cell and intra-cell interference as constructive elements, which help reduce the transmission power to achieve a target SINR. As comparison, although CoMP-Perfect-CSI scheme works as a network-level MISO to obtain a broadcast channel in multi-cell systems, the multiuser interference can not be utilized and needs to be canceled as much as possible. As a result, even with perfect CSI, CoMP-Perfect-CSI  is inferior to the proposed Full-CI-Prob/Det algorithms.
	For the Partial-CI-Prob and Partial-CI-Det algorithms, since the intended transmitted data is not shared among the BSs to reduce the coordination overhead, only intra-cell (multiuser) interference is utilized as constructive element while inter-cell interference is carefully suppressed by joint precoding design. As a result, the Partial-CI-Prob/Det algorithms consume higher power than the Full-CI-Prob/Det algorithms. However, compared to the most relevant CBF benchmarks in \cite{Nasseri2016Chance} and \cite{Shen2012Distributed}, the Partial-CI-Prob/Det algorithms consume much lower power, benefiting from utilizing multiuser interference as beneficial element. 
	At last, the Stat-CI algorithm only needs to share statistical inter-cell channel CSI with its adjacent cells. As a result, the inter-cell interference suppressing is not accurate as that in the instantaneous CSI based Partial-CI-Prob/Det algorithms. Hence, with the lightest complexity, the Stat-CI algorithm demonstrates a slight power consumption increment over the Partial-CI-Prob/Det algorithms. However, it is observed that the Stat-CI algorithm consumes significantly less power compared to the CBF benchmarks in \cite{Nasseri2016Chance} and \cite{Shen2012Distributed}, benefiting from utilizing intra-cell interference.

	Secondly, it is observed that the deterministic robust optimization generally requires more transmission power than the corresponding probabilistic robust optimization.  It is because the deterministic robust optimization needs to guarantee the SINR requirements satisfied all the time. Differently,  the probabilistic robust optimization satisfies the SINR requirement in a statistical way, where SINR outage is allowed to occur in a proper way. Hence, with a given threshold $\eta_{ik}$, the  probabilistic robust optimization consumes less power compared to its deterministic counterpart.
	Thirdly, the power consumption of all the algorithms increase with higher value of users' SINR requirement.

	\begin{figure}
		\centering
		\includegraphics[width=3.0 in]{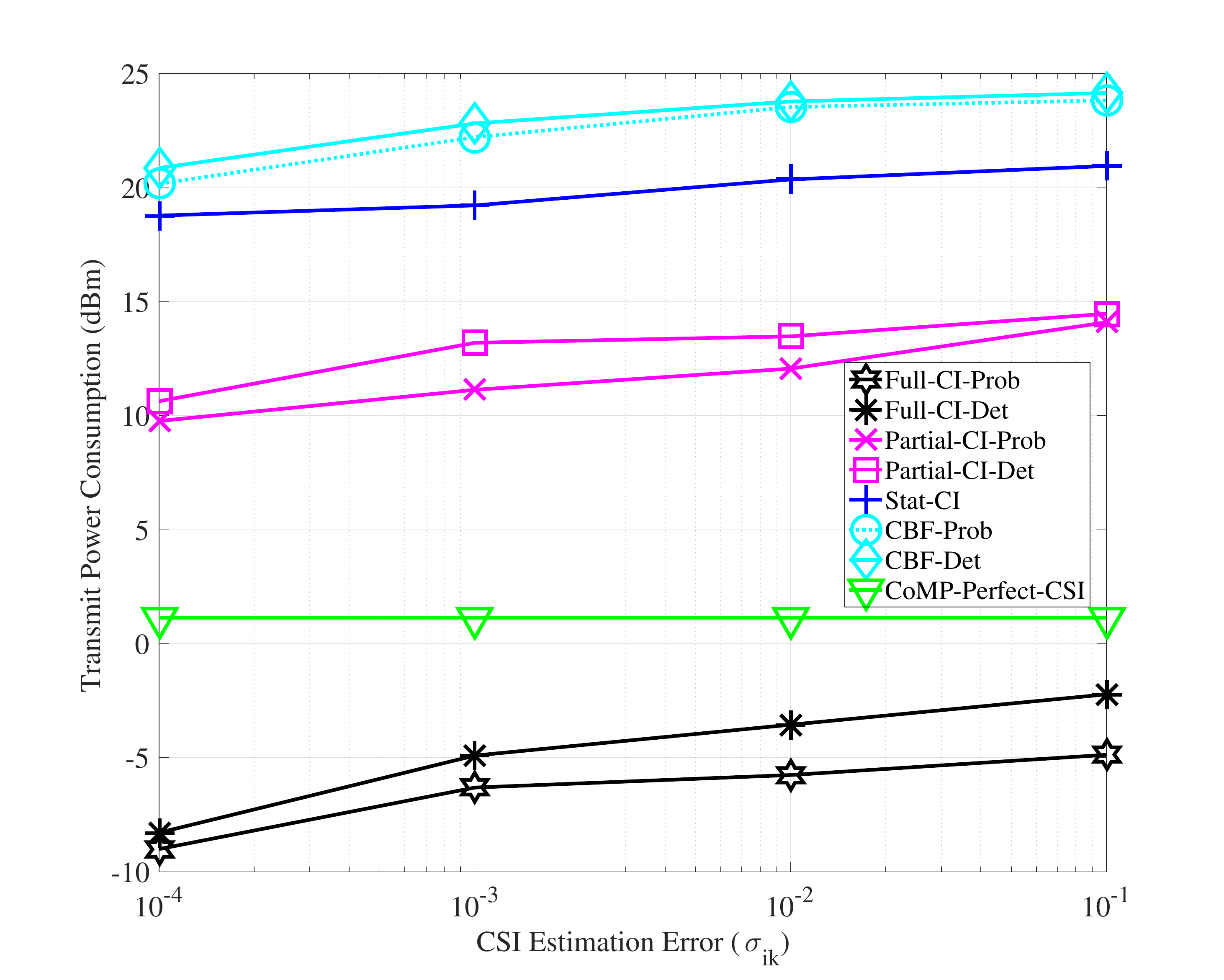}
		\caption{Impact of channel estimation error on the total power consumption, with $K=3$ in each cells and $\overline{\Gamma_{ik}}=20$ dB.}
		\label{fig:ErrorvsPower20181204}
	\end{figure}

	Fig. \ref{fig:ErrorvsPower20181204} shows how the channel estimation error affects the transmission power. For the probabilistic manner optimization, a tough channel estimation error increases the norm of covariance matrices, i.e.,  $\bm{\Theta}_{ik}$ and  $\bm{\Lambda}_{ik}$, of the estimated channels. According to Eqs. (\ref{eq:CI probability 4}) and (\ref{eq:CI probability 5}),  the amplitude of the precoder (also the transmission power) has to be improved to make the optimization feasible, leading to a increased transmission power. The same trend is also applicable for the Partial-CI-Prob algorithm.
	On the other hand, the deterministic manner robust optimization needs to keep the positive semi-definite characteristic for the matrices in Eqs. \eqref{eq:CI det SINR 7} and  \eqref{eq:CI det SINR 8}, which mathematically requires that all the leading principal minors in the matrices to be nonnegative.  As a result, the transmission power of all the deterministic robust optimization increases with a tough CSI estimation. 
	At last, since the CoMP-imperfect scheme is designed based on the ideal CSI acquisition, its power consumption is independent with the CSI error.

	\begin{figure}
		\centering
		\includegraphics[width=3.0 in]{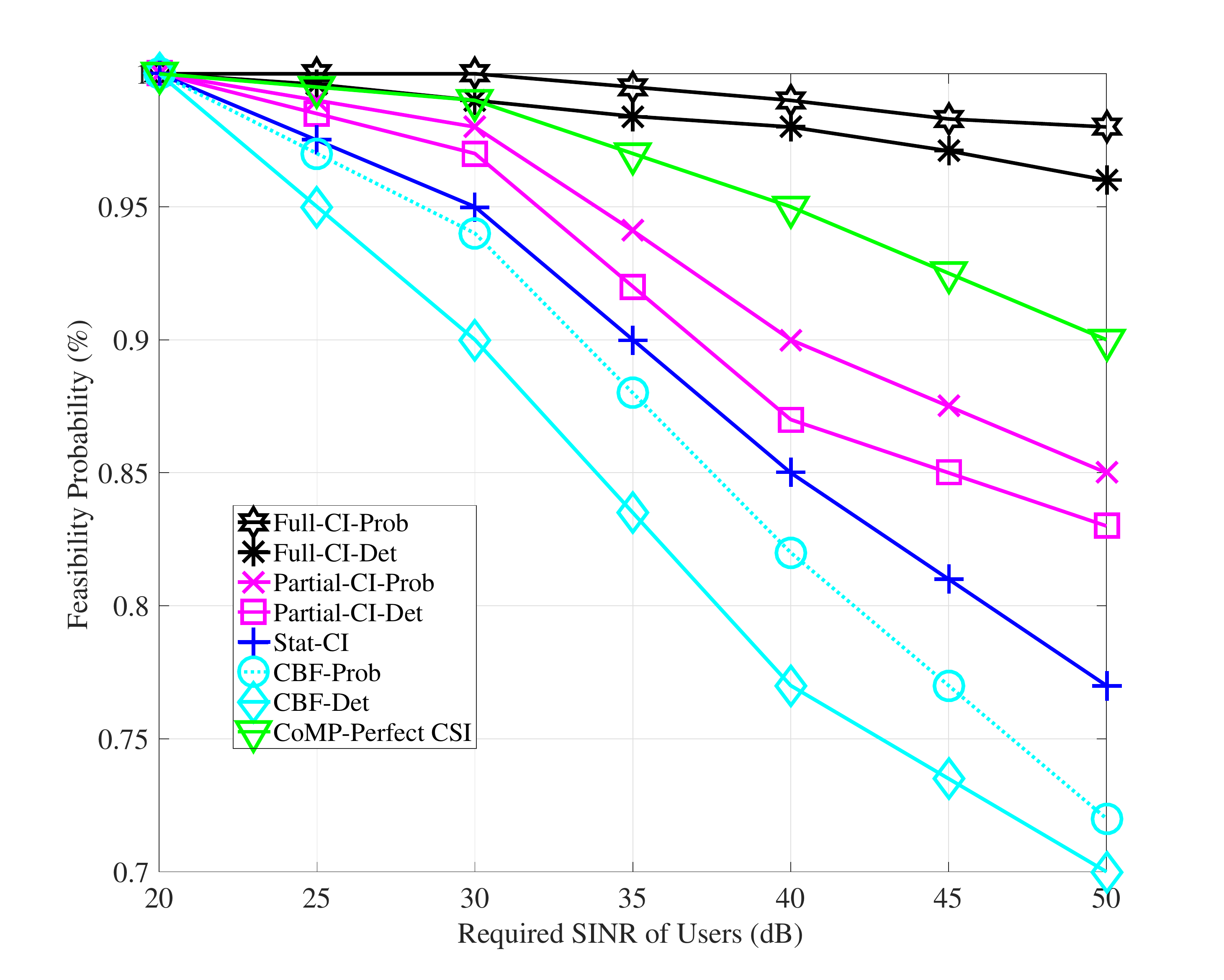}
		\caption{Feasibility probability (\%) versus different SINR requirements $\overline{\Gamma_{ik}}$, with $K=3$ in each cells and $\sigma_{ik}=10^{-3}$.}
		\label{fig:SINRvsFeasibility20181204}
	\end{figure}
	
	Fig. \ref{fig:SINRvsFeasibility20181204} shows the feasibility probability with different value of SINR requirements, where infeasibility occurs when the consumed power violates the maximum available power constraints $p_{max}$. As can be seen, the Full-CI-prob and Full-CI-det achieves the highest feasibility probabilities, up to 95\% even with SINR requirement $\Gamma_{ik}=50$ dB. As a comparison, the feasibility probability of the CoMP-perfect CSI degrades to 90\% with $\Gamma_{ik}=50$ dB SINR requirement. 
	Besides, the Partial-CI-Prob/Det algorithms achieve lower feasibility probabilities compared to the CoMP-like schemes, as inter-channel gain is considered as harmful element and is suppressed. As a result, they have higher probabilities of violating maximum power budget to achieve a target SINR, e.g., around 85\% feasibility probability with $\Gamma_{ik}=50$ SINR requirement. At last, neither inter-cell or intra-cell interference is exploited by the conventional  CBF-Prob and CBF-Det benchmarks, and hence their feasibility probabilities sharply degrade to 70\%, lower than the 77\% achieved by the Stat-CI algorithm.

	\begin{figure}
		\centering
		\includegraphics[width=3.0 in]{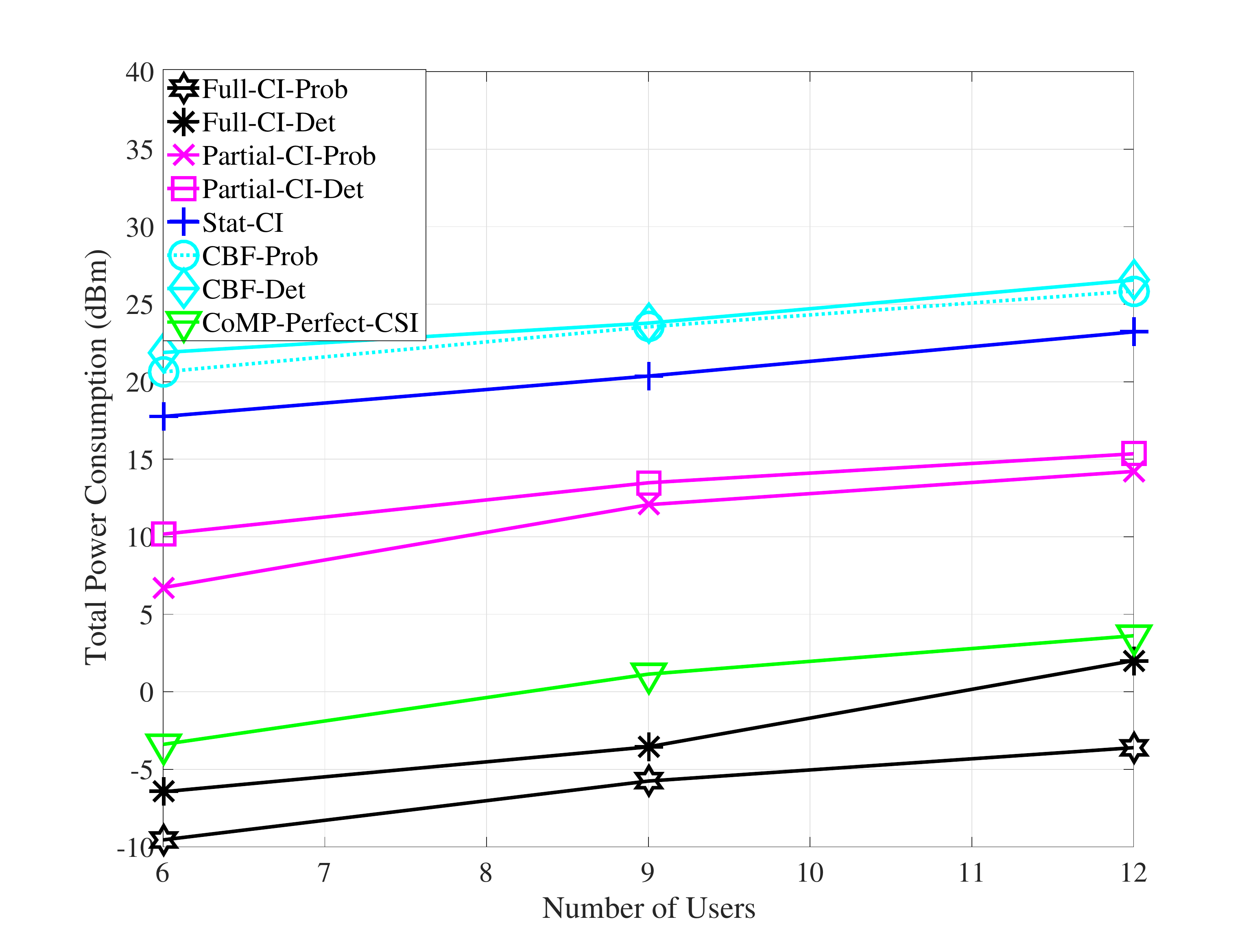}
		\caption{Impact of the total number of users on the total power consumption, with $\overline{\Gamma_{ik}}=20$ dB and $\sigma_{ik}=10^{-2}$.}
		\label{fig:UservsPower20181121}
	\end{figure}
	Fig. \ref{fig:UservsPower20181121} demonstrates the power consumption versus different number of users. As can be seen, the proposed algorithms consume less power compared to their corresponding benchmarks with different number of multiple users. Besides, higher power consumption is required with more users. It is because according to the problem formulations, all the users' SINR requirements need to be  probabilistically or deterministically satisfied by the constraints. As a result, higher transmission power is led due to the increased number of constraints.

	\begin{figure}
		\centering
		\includegraphics[width=4.5 in]{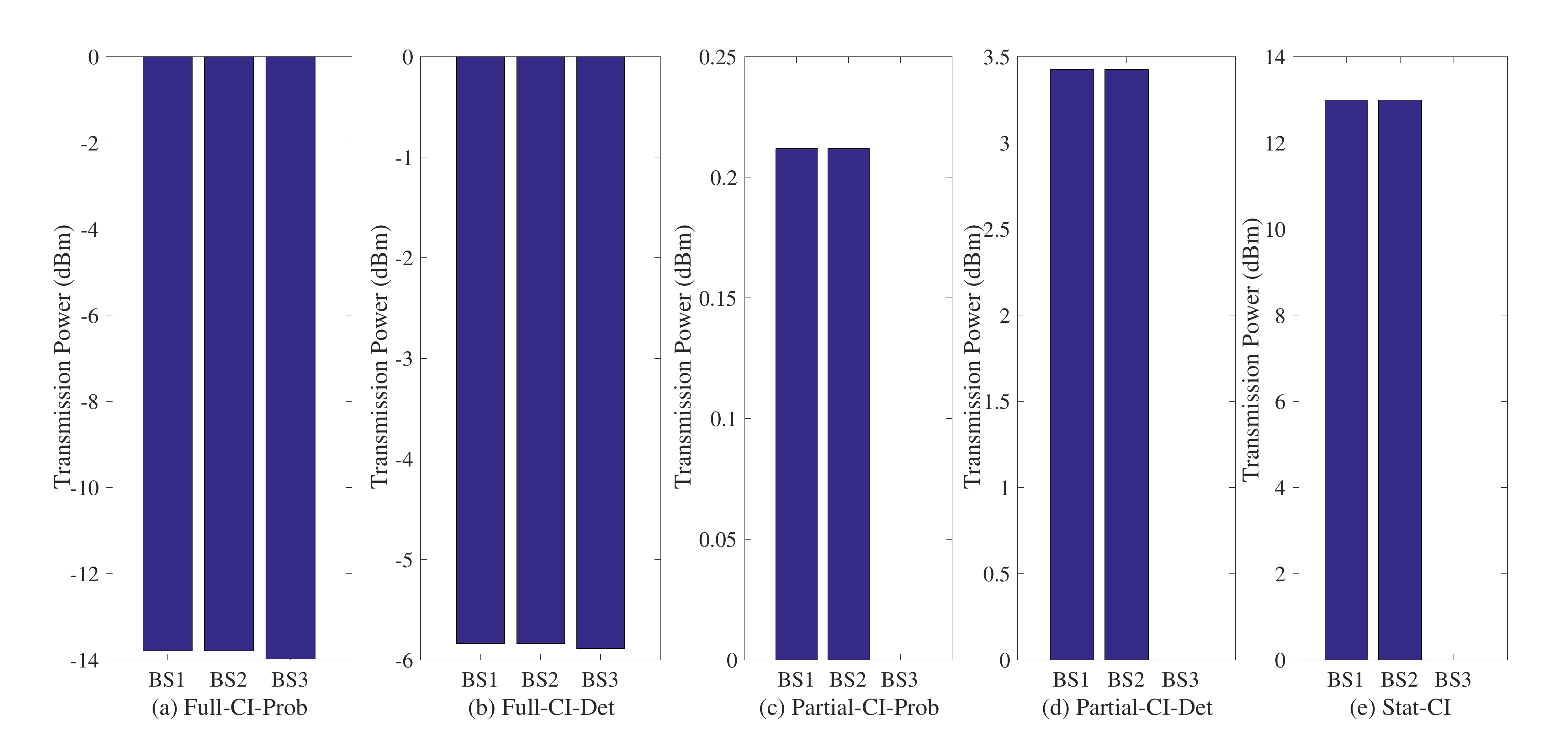}
		\caption{Transmission power of the BSs under different algorithms, with $\overline{\Gamma_{ik}}=20$ dB and  $\sigma_{ik}=10^{-2}$.}
		\label{fig:Cooperation20181121}
	\end{figure}
	Fig. \ref{fig:Cooperation20181121} shows the transmission power for the different level of coordination algorithms. For illustration, we place 6 users in the BS1 and BS2's edge area while no user in the BS3's coverage. As can be seen, by the Full-CI-Prob/Det algorithms, although there is no user in the coverage area of the BS3, the BS3 still coordinates with the BS1 and BS2 as a network-level MISO to serve these edge users. Hence, as presented by Figs. \ref{fig:Cooperation20181121} (a) and (b), the BS3 contributes almost identical transmission power compared to the BS1 and BS2. In contrast, by the Partial-CI-Prob/Det algorithms, the BS3 keeps silent when there is no user within its cell, as shown by  Figs. \ref{fig:Cooperation20181121} (c) and (d). It is because by the Partial-CI-Prob/Det algorithms, BSs only transmits data to the user within its cell, while the inter-cell gain in strictly suppressed by joint precoding design. The same trend is observed by the Stat-CI scheme, which suppresses inter-cell interference with the knowledge of statistical inter-cell channels.

	\begin{figure}
		\centering
		\includegraphics[width=6.3 in]{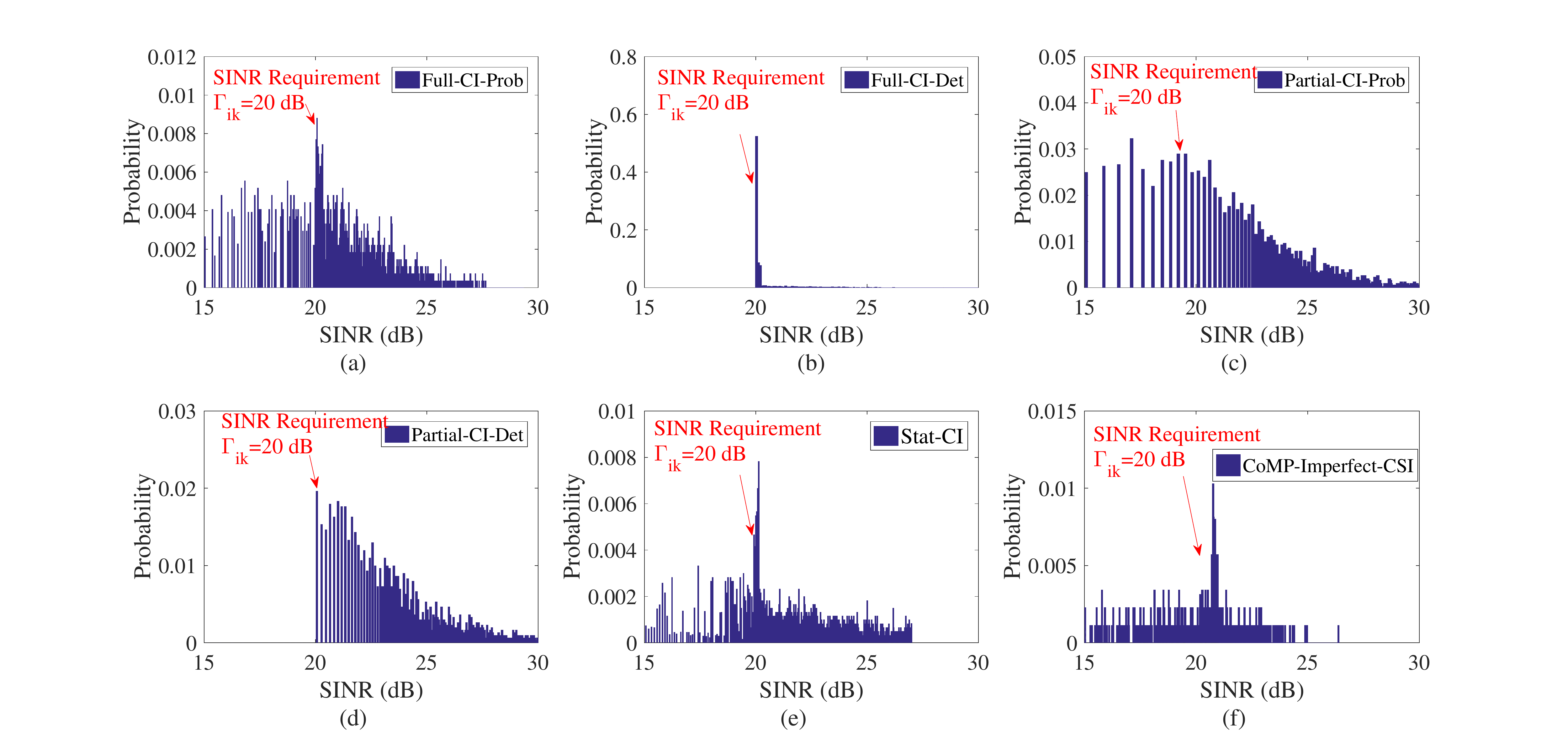}
		\caption{Probability distribution function (pdf) of the SINRs by the different algorithms a) Full-CI-Prob, b) Full-CI-Det, c) Partial-CI-Prob, d) Partial-CI-Det, e) Stat-CI, f) Conventional CoMP (Imperfect CSI case), where $\eta_{ik}=80\%$. SINR requirement $\overline{\Gamma_{ik}}=20$ dB).}
		\label{fig:distribution20181212}
	\end{figure}

	Fig. \ref{fig:distribution20181212} shows the distribution of SINRs obtained by the different algorithms.
	Firstly, it can be seen from Figs. \ref{fig:distribution20181212} (b) and (d) that, all the worst-case based optimizations, i.e., Full-CI-Det and Partial-CI-Det, can always guarantee that the achieved SINR higher than the preset SINR requirements. In contrast, the chance-constrained based optimizations, i.e., Full-CI-Prob and Partial-CI-Prob in Figs. \ref{fig:distribution20181212} (a) and (c), allow proper outage occurs obeying the preset threshold $\eta_{ik}$.
	Secondly, by comparing Full-CI and Partial-CI schemes, it is observed that the achieved SINR by the Partial-CI scheme is more long-tailed, e.g., from 25 dB to 30 dB in Figs. \ref{fig:distribution20181212} (c) and (d).
	It is because a BS may radiate high transmit power to compensate the edge users' SINR by the Partial-CI scheme, and hence the users close to the BS may occasionally obtain extreme high SINR, leading to a high pdf value at high SINR regime. Different, the Full-CI scheme efficiently corporates adjacent BSs to serve edge users without radiating high transmission power, and hence all the users' SINR is less expanded and centers on the SINR target.  
	Thirdly, it is observed that the SINR performance of the Stat-CI algorithm is impaired due to only knowing the statistical inter-cell channel. However, by taking into account local imperfect CSI, the Stat-CI still outperforms the conventional CoMP-Imperfect-CSI.

	\begin{figure}
		\centering
		\includegraphics[width=3.0 in]{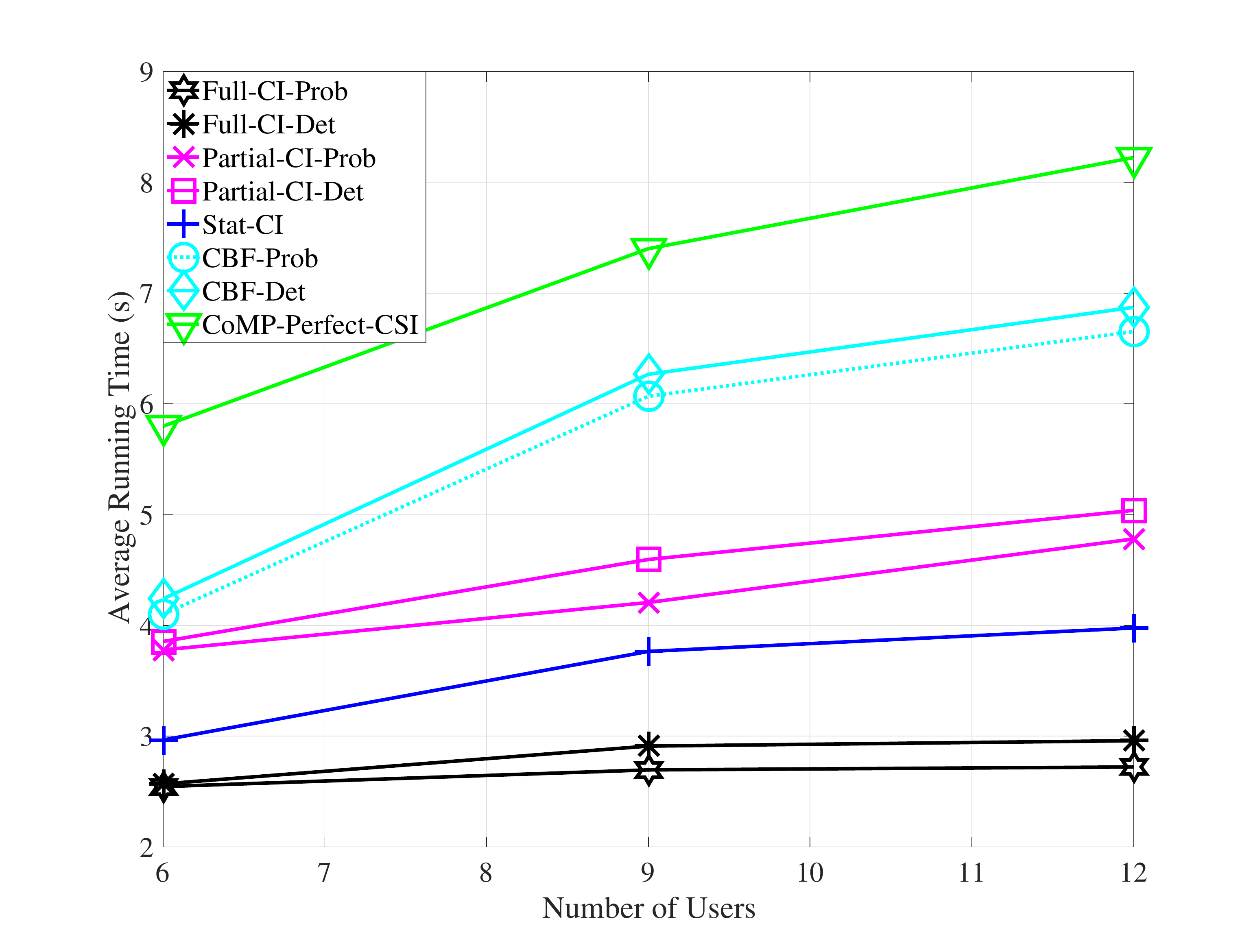}
		\caption{Impact of the total number of users on the execution time, with $\overline{\Gamma_{ik}}=20$ dB and  $\sigma_{ik}=10^{-2}$.}
		\label{fig:Runningtime20181121}
	\end{figure}
	Fig. \ref{fig:Runningtime20181121} shows the average running time of different algorithms versus the number of users. It can be seen proposed algorithms require a lower running time. 
	Among them, the Partial-CI-Prob/Det algorithms require the highest time for obtaining the optimal results. It corresponds to our analytic analysis that the Partial-CI-Prob/Det algorithms are confined by more constraints, and more slack variables are introduced to make the problem solvable. Hence, the Partial-CI-Prob/Det algorithms require longer time to get the optimal results.
	Besides, all the three benchmarks, CoMP-Perfect-CSI  and CBF-Prob/Det, need longer time to get convergence, although they subject to fewer constraints. It is because by the conventional methods, the number of variables (precoders)  exactly equals to the number of users. In contrast, the proposed schemes transform the transmission channel into a multi-cast channel by utilizing CI, and hence the number of variables only equals to the number of the coordinated BSs. As a result, even though the proposed algorithms subject to more constraints, a shorter running time  is provisioned over the conventional schemes. 
	With in-total 12 users served by $N_{BS}=3$ BSs, there are 12 variables in the conventional algorithms while the number of variables is reduced to 3 by the proposed CI-based algorithms. As a result, the algorithms require 2-5 seconds less than the conventional algorithms.

	\begin{figure}
		\centering
		\includegraphics[width=3.0 in]{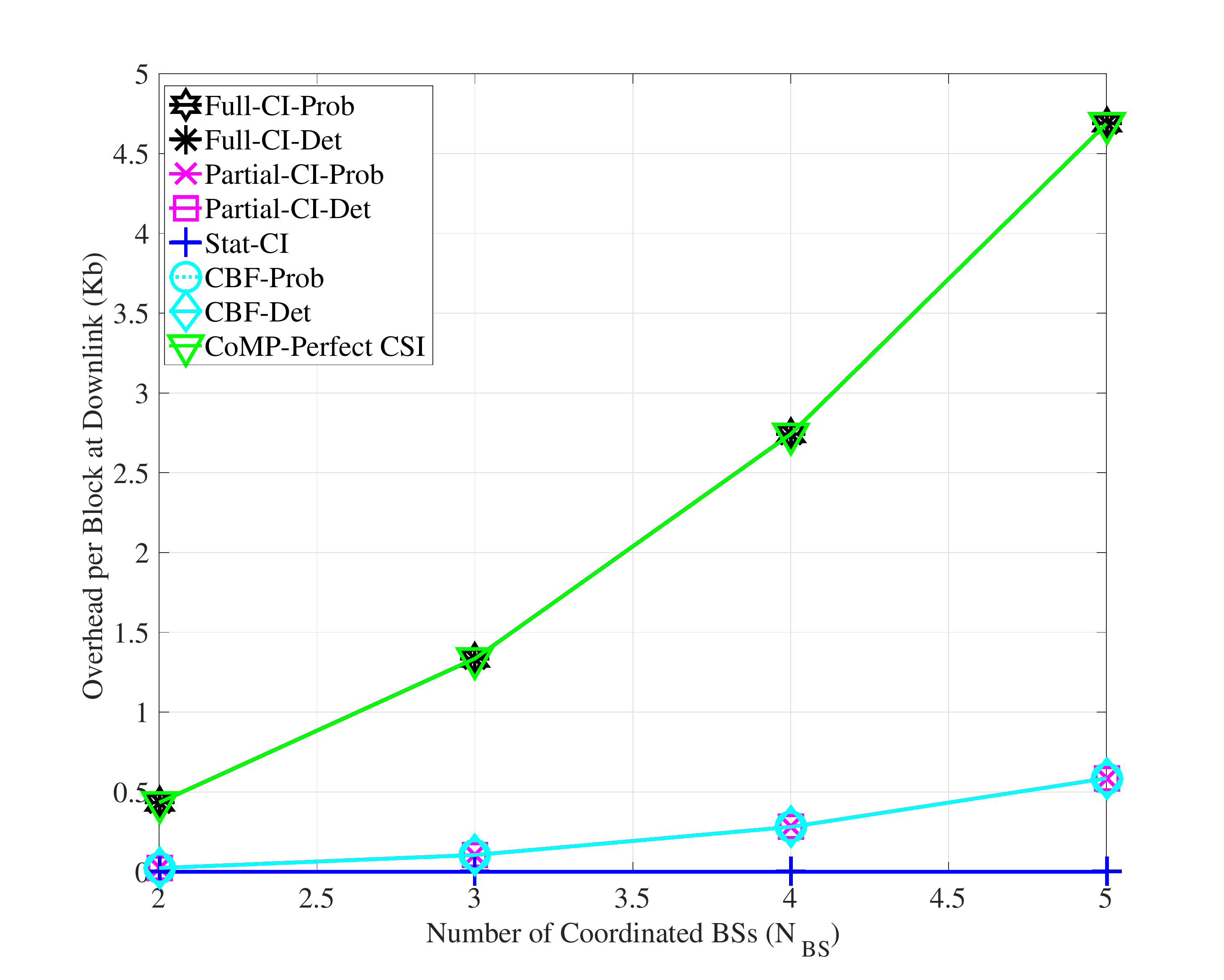}
		\caption{The total coordination overhead in each sub-frame versus the number of the coordinated BSs.}
		\label{fig:Overhead20181121}
	\end{figure}

	Fig. \ref{fig:Overhead20181121} demonstrates the total coordination overhead required by the different schemes. As analyzed in Section VI, the total coordination overhead consists of sharing CSI and symbols.  For CSI exchange, codebook-based CSI overhead is accounted since it requires less bits to describing CSI over the analog-feedback method.
	Explicitly, by the codebook-based CSI acquisition, BSs only need to  share the indexes of the selected codeword with other BSs, and then each BS locally interprets the codewords which are closest to the projection of the CSI. Generally, around 2-10 bits are needed to describe a MISO channel for each user \cite{Huang2018A}.
	For the overhead on sharing symbols, LTE Type 2 downlink frame structure is considered, where 5 out of 10 sub-frames are designed for downlink transmission and each sub-frame contains 14 symbols.
	It can be seen, all the CoMP-like schemes, i.e, Full-CI-Prob/Det and CoMP-Perfect-CSI, need high overhead, since both the intended transmitted symbols and CSI have to be shared among BSs, and especially sharing the data introduces high coordination overhead.
	In the lower-level of coordination, Partial-CI-Prob/Det and CBF-Prob/Det require relatively moderate overhead, where only CSI needs to be shared among BSs. As a result, coordination overhead is reduced to 0.5 Kb when 5 BSs are coordinated.
	At last, it should be noted that the Stat-CI requires no backhaul overhead for coordination, since only statistical inter-cell CSI needs to be exchanged among the coordinated BSs, which can be ignored compared to other instantaneous CSI acquisition based algorithms.


	\section{CONCLUSIONS}

	In this paper, we have investigated robust power-efficient precoding design for multi-cell coordination systems, based on exploiting inter-cell and intra-cell interference. Three schemes have been proposed to fully/partially utilize inter-cell and intra-cell interference, with different levels of coordination overhead. At the presence of CSI estimation error, we have processed the power minimization problems in terms of probabilistic and deterministic robust optimizations. Explicitly, probabilistic robust optimization satisfies the SINR requirements in a statistical manner while deterministic robust optimization satisfies the SINR requirement all the time.
	The simulation results verify that, by sharing the intended transmitted symbols and CSI, the Full-CI-Prob/Det algorithms consume the lowest power, and even outperform the conventional CoMP with perfect CSI acquisition. 
	Differently, by only sharing CSI but not the intended transmitted symbols, the Partial-CI-Prob/Det algorithms require moderate coordination overhead whereas outperforming the conventional  CBF-based schemes in terms of power efficient transmission. 
	To further reduce coordination overhead, the finally proposed Stat-CI algorithm obtains reasonable transmission power consumption with the lightest overhead.
	Last but not least, the proposed algorithms require low running time, benefiting from the multi-cast transmission characteristics of CI precoding.

	\begin{appendices}
		
		\section{Proof of Lemma 2}
		
		Since the element of CSI error vector $\bm{e}_{ik}$ follows normal distribution such that $\mathcal{CN}(0,\sigma_{ik}^2)$, the element of real part $\bm{e}_{ik}^{\Re}$ and imaginary part $\bm{e}_{ik}^{\Im}$ follows $\mathcal{CN}(0, \frac{ \sigma_{ik}^2}{2})$. Hence, $(\sqrt{\bm{c}_{ik}}) ^T\bm{I}_{2M\times N_{BS}} \sqrt{\bm{c}_{ik}}  - \xi^2 \leq 0$ can be equivalently written as
		$  \sum_{j=1}^{N_{BS}} \big(  (1+\mathrm{tan}\theta) \bm{e}_{ik}^{\Im} +( 1-\mathrm{tan}\theta )\bm{e}_{ik}^{\Re}  \big)\leq \xi^2    $, which can be approximately seen as the probability distribution function (pdf) of a normal distributed variable such that $\mathrm{Pr}\{ \sum_{j=1}^{N_{BS}} \big(  (1+\mathrm{tan}\theta) \bm{e}_{ik}^{\Im} +( 1-\mathrm{tan}\theta )\bm{e}_{ik}^{\Re}  \big)\leq \xi^2 \} = \delta$, where $\delta$ physically represents the probability of inequality being satisfied.
		The value of $\delta$ can be set to close to 1, i.e. $\delta=0.99$, meaning the inequality is satisfied with a high probability.  The term  $ \sum_{j=1}^{N_{BS}} \big(  (1+\mathrm{tan}\theta) \bm{e}_{ik}^{\Im} +( 1-\mathrm{tan}\theta )\bm{e}_{ik}^{\Re}\big)$  follows $\mathcal{CN}(0, MN_{BS}\sigma_{ik}^2(1+ \mathrm{tan}^2\theta  ))$. By normalizing it into a standard normal distributed variable, we get a cumulative distribution function (cdf) as $\Phi \{   \frac{\xi^2}{ \sigma_{ik} \sqrt{  MN_{BS}(1+ \mathrm{tan}^2\theta ) }   }   \} = \delta$, where $\Phi(\cdot)$ is the cdf of a standard normal distributed variable. Defining $\Phi^{-1}(\cdot)$ as the inverse function of $\Phi(\cdot)$, finally we get $\xi^2 =\Phi^{-1}(   \delta )\sigma_{ik} \sqrt{  MN_{BS}(1+ \mathrm{tan}^2\theta ) }$.

		


		\section{Proof of Lemma 3}
		
		The term $\bm{e}_{ik}^H\bm{I}\bm{e}_{ik}  \leq \nu^2 $ can be equivalently written as  $ (\frac{[\bm{e}_{ik}]_{1}}{\sigma_{ik}})^2 + ...+(\frac{[\bm{e}_{ik}]_{M}}{\sigma_{ik}})^2   \leq (\frac{\nu}{\sigma_{ik}})^2 $. We notice its left hand follows chi-square distribution with $M$ degrees of freedom. Hence, $\bm{e}_{ik}^H\bm{I}\bm{e}_{ik}  \leq \nu^2 $ can be approximately interpreted as the pdf of a chi-square distributed variable with $M$ degrees of freedom such that $\mathrm{Pr}\{ (\frac{[\bm{e}_{ik}]_{1}}{\sigma_{ik}})^2 + ...+(\frac{[\bm{e}_{ik}]_{M}}{\sigma_{ik}})^2      \leq \frac{\nu^2}{\sigma_{ik}^2}  \} = \delta$, where $\delta$ physically represents the probability of the inequality being satisfied.
		The value of $\delta$ can be set to close to 1, i.e. $\delta=0.99$. Evidently, it can be written as a cdf as $\Upsilon \{   \frac{\nu^2}{ \sigma_{ik} ^2    }   \} = \delta$, where $\Upsilon(\cdot)$ is the cdf of a chi-square distributed variable. Defining $\Upsilon^{-1}(\cdot)$ as the inverse function of $\Phi(\cdot)$, finally we get $\xi^2 =\Upsilon^{-1}(   \delta )\sigma_{ik}^2.$



	\end{appendices}

	\bibliographystyle{IEEEbib}
	\bibliography{CImulticell20181212}

\end{document}